\documentclass[twocolumn,longauth]{aa}
\usepackage{graphicx}
\usepackage{color}
\usepackage{ulem}
\usepackage{txfonts}
\usepackage{xspace}

\usepackage{url}
\usepackage{lineno}

\newcommand{\pks}{\object{PKS\,2155$-$304}\xspace}
\newcommand{\pg}{\object{PG\,1553$+$113}\xspace}
\newcommand{\hess}{H.E.S.S.\xspace}
\newcommand{\fermi}{\textit{Fermi}-LAT\xspace}
\newcommand{\phasetwo}{\hess~II\xspace}
\newcommand{\fvar}{\mathrm{F}_{\mathrm{var}}}

\usepackage{natbib,twoopt}
\usepackage[breaklinks=true]{hyperref} 
\bibpunct{(}{)}{;}{a}{}{,}             
\makeatletter
  \newcommandtwoopt{\citeads}[3][][]{\href{http://adsabs.harvard.edu/abs/#3}%
    {\def\hyper@linkstart##1##2{}%
     \let\hyper@linkend\@empty\citealp[#1][#2]{#3}}}
  \newcommandtwoopt{\citepads}[3][][]{\href{http://adsabs.harvard.edu/abs/#3}%
    {\def\hyper@linkstart##1##2{}%
     \let\hyper@linkend\@empty\citep[#1][#2]{#3}}}
  \newcommandtwoopt{\citetads}[3][][]{\href{http://adsabs.harvard.edu/abs/#3}%
    {\def\hyper@linkstart##1##2{}%
     \let\hyper@linkend\@empty\citet[#1][#2]{#3}}}
  \newcommandtwoopt{\citeyearads}[3][][]%
    {\href{http://adsabs.harvard.edu/abs/#3}
    {\def\hyper@linkstart##1##2{}%
     \let\hyper@linkend\@empty\citeyear[#1][#2]{#3}}}
    \makeatother
    
\begin{document}

\title{Gamma-Ray Blazar Spectra with \phasetwo Mono Analysis: The Case of \pks and \pg}
\authorrunning{H.E.S.S.~Collaboration \& LAT~Collaboration}
\titlerunning{Gamma-Ray Blazar Spectra with \phasetwo Mono Analysis}
\author{\small
\hess~Collaboration
\and H.~Abdalla \inst{1}
\and A.~Abramowski \inst{2}
\and F.~Aharonian \inst{3,4,5}
\and F.~Ait Benkhali \inst{3}
\and A.G.~Akhperjanian\protect${}^\dagger$ \inst{6,5} 
\and T.~Andersson \inst{10}
\and E.O.~Ang\"uner \inst{7}
\and M.~Arrieta \inst{15}
\and P.~Aubert \inst{24}
\and M.~Backes \inst{8}
% ...
}

\institute{
Centre for Space Research, North-West University, Potchefstroom 2520, South Africa \and 
Universit\"at Hamburg, Institut f\"ur Experimentalphysik, Luruper Chaussee 149, D 22761 Hamburg, Germany \and 
Max-Planck-Institut f\"ur Kernphysik, P.O. Box 103980, D 69029 Heidelberg, Germany \and 
Dublin Institute for Advanced Studies, 31 Fitzwilliam Place, Dublin 2, Ireland \and 
National Academy of Sciences of the Republic of Armenia,  Marshall Baghramian Avenue, 24, 0019 Yerevan, Republic of Armenia  \and
Yerevan Physics Institute, 2 Alikhanian Brothers St., 375036 Yerevan, Armenia \and
Institut f\"ur Physik, Humboldt-Universit\"at zu Berlin, Newtonstr. 15, D 12489 Berlin, Germany \and
University of Namibia, Department of Physics, Private Bag 13301, Windhoek, Namibia \and
GRAPPA, Anton Pannekoek Institute for Astronomy, University of Amsterdam,  Science Park 904, 1098 XH Amsterdam, The Netherlands \and
Department of Physics and Electrical Engineering, Linnaeus University,  351 95 V\"axj\"o, Sweden \and
Institut f\"ur Theoretische Physik, Lehrstuhl IV: Weltraum und Astrophysik, Ruhr-Universit\"at Bochum, D 44780 Bochum, Germany \and
GRAPPA, Anton Pannekoek Institute for Astronomy and Institute of High-Energy Physics, University of Amsterdam,  Science Park 904, 1098 XH Amsterdam, The Netherlands \and
Institut f\"ur Astro- und Teilchenphysik, Leopold-Franzens-Universit\"at Innsbruck, A-6020 Innsbruck, Austria \and
School of Physical Sciences, University of Adelaide, Adelaide 5005, Australia \and
LUTH, Observatoire de Paris, PSL Research University, CNRS, Universit\'e Paris Diderot, 5 Place Jules Janssen, 92190 Meudon, France \and
Sorbonne Universit\'es, UPMC Universit\'e Paris 06, Universit\'e Paris Diderot, Sorbonne Paris Cit\'e, CNRS, Laboratoire de Physique Nucl\'eaire et de Hautes Energies (LPNHE), 4 place Jussieu, F-75252, Paris Cedex 5, France \and
DSM/Irfu, CEA Saclay, F-91191 Gif-Sur-Yvette Cedex, France \and
Astronomical Observatory, The University of Warsaw, Al. Ujazdowskie 4, 00-478 Warsaw, Poland \and
Aix Marseille Universit\'e, CNRS/IN2P3, CPPM UMR 7346,  13288 Marseille, France \and
Instytut Fizyki J\c{a}drowej PAN, ul. Radzikowskiego 152, 31-342 Krak{\'o}w, Poland \and
Funded by EU FP7 Marie Curie, grant agreement No. PIEF-GA-2012-332350,  \and
Laboratoire Univers et Particules de Montpellier, Universit\'e Montpellier, CNRS/IN2P3,  CC 72, Place Eug\`ene Bataillon, F-34095 Montpellier Cedex 5, France \and
School of Physics, University of the Witwatersrand, 1 Jan Smuts Avenue, Braamfontein, Johannesburg, 2050 South Africa \and
Laboratoire d'Annecy-le-Vieux de Physique des Particules, Universit\'{e} Savoie Mont-Blanc, CNRS/IN2P3, F-74941 Annecy-le-Vieux, France
% ...
}

\offprints{H.E.S.S. and LAT~collaborations,
\protect\\\email{\href{mailto:contact.hess@hess-experiment.eu}{contact.hess@hess-experiment.eu}};
\protect\\${}^*$ Corresponding Authors
\protect\\${}^\dagger$ Deceased}

\date{Received XXX; accepted XXX}

\abstract{The addition of a 28~m Cherenkov telescope (CT5) to the 
\hess array extended the experiment's sensitivity to lower energies.
The lowest energy threshold is obtained using monoscopic analysis of data taken with 
CT5, providing access to gamma-ray energies below 100~GeV for small zenith
angle observations. Such an extension of the
instrument's energy range is particularly beneficial for studies of Active Galactic 
Nuclei (AGNs) with soft spectra, as expected for those at a redshift $\geq$ 0.5.
The high-frequency peaked BL Lac objects \pks ($z=0.116$) and \pg ($0.43 <z<0.58$)
are among the brightest objects in the gamma-ray sky,
both showing clear signatures of gamma-ray absorption at $E > 100$~GeV
interpreted as being due to interactions with the extragalactic background light (EBL).
}
{This work is directed toward a twofold aim:
to demonstrate the monoscopic analysis of CT5 data with a low energy threshold,
and to obtain accurate measurements of the spectral energy distributions (SED)
of \pks and \pg near their SED peaks at energies $\approx$ 100~GeV.}
{Multiple observational campaigns of \pks and \pg were conducted
during 2013 and 2014 using the full \phasetwo instrument (CT1--5).
A monoscopic analysis of the data taken with the new CT5 telescope was developed along
with an investigation into the systematic uncertainties on the spectral
parameters which are derived from this analysis.}
{Using the data from CT5, the energy spectra of \pks and \pg were reconstructed 
down to conservative threshold energies of 80~GeV for \pks, which transits near zenith, and 110~GeV for the more 
northern \pg. The measured spectra, well fitted in both cases by a log-parabola spectral 
model (with a 5.0~$\sigma$ statistical preference for non-zero curvature for \pks and 
4.5~$\sigma$ for \pg), were found consistent with spectra derived from contemporaneous 
\fermi data, indicating a sharp break in the observed spectra of both sources at 
$E \approx 100$~GeV. When corrected for EBL absorption, 
the intrinsic \phasetwo mono and \fermi spectrum of \pks was 
found to show significant curvature. For \pg, however, no significant detection of 
curvature in the intrinsic spectrum could be found within statistical and systematic 
uncertainties.}{}


\maketitle

\section{Introduction}
\label{intro}

The very high energy (VHE, $E\gtrsim 100$~GeV) gamma-ray experiment of 
the High Energy Stereoscopic System (\hess) consists of five imaging 
atmospheric Cherenkov telescopes (IACTs) located in the Khomas Highland 
of Namibia ($23\degr 16\arcmin 18\arcsec$ S, $16\degr 30\arcmin 
00\arcsec$ E), 1835~m above sea level. From January 2004 to October 
2012, the array was operated as a four telescope instrument (\hess 
phase I). The telescopes, CT1--4, are arranged in a square formation 
with a side length of 120 m. Each of these telescopes has an effective 
mirror surface area of 107~m$^{2}$, a field of view of  $5\degr$ in 
diameter, capable of detecting cosmic gamma rays in the energy range 
0.1--100~TeV \citepads{2006A&A...457..899A}. In October 2012 a fifth 
telescope, CT5, placed at the centre of the original square, started 
taking data. This set-up is referred to as \hess phase II, or 
\phasetwo. With its effective mirror surface close to 600~m$^{2}$ and a 
fast, finely pixelated camera \citepads{2014NIMPA.761...46B}, CT5 
potentially extends the energy range covered by the array down to 
energies of $\sim 30$~GeV.

In this study, we focus on obtaining high statistic results with observations of the 
high-frequency peaked BL Lac objects \pks and \pg. These blazars are among the 
brightest objects in the VHE gamma-ray sky. Furthermore, the spectra of both these
blazars exhibit  signatures of gamma-ray absorption at energies $E \sim 100$~GeV,
due to interactions with the extragalactic background light (EBL).

 \pks is a high-frequency peaked BL Lac 
(HBL) object at $z=0.116$ \citepads{2013MNRAS.435.1233G}. This source 
is located in a galaxy poor cluster \citepads{1993ApJ...411L..63F} and 
the host galaxy is resolved \citepads{1998A&A...336..479K}. It was 
first discovered as a high energy emitter by the HEAO 1 X-ray satellite 
\citepads{1979ApJ...234..810G, 1979ApJ...229L..53S}. Gamma-ray emission 
in the energy range 30~MeV to 10~GeV was detected from this blazar by 
the EGRET instrument on board the Compton Gamma Ray Observatory 
\citepads{1995IAUC.6169....1V}. The first detection in the VHE range 
was attained in 1996 by the University of Durham Mark 6 Telescope, with 
a statistical significance of $6.8~\sigma$ 
\citepads{1999ApJ...513..161C}. Starting from 2002 the source was 
regularly observed with \hess, with the first detection based on the 
2002 data subsequently published with just one telescope of \hess phase 
I \citepads{2005A&A...430..865A}. After completion of the array, this 
source was detected in stereoscopic mode in 2003 with high significance 
($> 100~\sigma$) at energies greater than 160 GeV 
\citepads{2005A&A...430..865A}. Strong flux variability with multiple 
episodes of extreme flaring activity in the VHE band were reported 
\citepads{2007ApJ...664L..71A,2010A&A...520A..83H,2012A&A...544A..75A}. 
A photon index \footnote{	 Where the photon index, $\Gamma$, describes
the spectral shape of the photon energy distribution, $dN/dE\propto E^{-\Gamma}$.} 
of $3.53 \pm 0.06_{\rm{stat}} \pm 0.10_{\rm{syst}}$ was obtained from 
analysis of observations during a low flux state (2005--2007) above 
200~GeV \citepads{2010A&A...520A..83H}. For average and high flux 
states the presence of curvature or a cut-off was favoured from the 
spectral fit analysis carried out \citepads{2010A&A...520A..83H}.


The HBL object \pg was first announced as a VHE gamma-ray source by 
\hess \citepads{2006A&A...448L..19A} and independently and almost 
simultaneously confirmed by MAGIC using observations from 2005 
\citepads{2007ApJ...654L.119A}. The \hess~I measurements 
\citepads{2008A&A...477..481A} yielded a photon index $\Gamma = 4.5 \pm 
0.3_{\rm{stat}} \pm 0.1_{\rm{syst}}$ above 225~GeV. At high energies 
(HE, 100~MeV $< E <$ 300~GeV) the source was detected by \fermi with a 
photon index of $1.68 \pm 0.03$ 
\citepads{2009ApJS..183...46A,2010ApJ...708.1310A}, making \pg an 
active galactic nucleus (AGN) with one of the largest HE-VHE spectral 
breaks observed and a hint for long-term gamma-ray flux oscillation \citep{2015ApJ...813L..41A}. The redshift of \pg is constrained by UV observations 
to the range $0.43 < z \lesssim 0.58$ \citepads{2010ApJ...720..976D}. 
The first upper-limits of $z<0.69$ (pre-\fermi) \citetads{2007ApJ...655L..13M} 
and more recently (post-\fermi) $z<0.61$ on the source redshift have been 
obtained \citetads{2015ApJ...799....7A} using TeV data 
and of $z < 0.53$ by \citetads{2015ApJ...812...60B} using also GeV data. 
Assuming that the difference in spectral indices between the HE and VHE regimes is 
imprinted by the attenuation by the extragalactic background light, the 
redshift was constrained to the range $z = 0.49 \pm 0.04$ 
\citepads{2015ApJ...802...65A}.

This paper reports on the first observations of \pks and \pg conducted in 2013 and 2014 
using the \phasetwo instrument (CT5) in monoscopic mode. A description of the analysis for both AGNs, using data 
from this instrument, is provided. Systematic errors associated with our 
results are also estimated. Particular emphasis is placed on the spectral measurements at low 
energies and their connection with the \fermi measurements. Using the
\phasetwo mono and \fermi results, the implications on intrinsic source spectrum are 
considered.

\section{The \phasetwo experiment}

The \phasetwo experiment is the first hybrid Cherenkov instrument and has the
ability to take data in different modes. The \phasetwo system triggers on events
detected either by CT5 only (mono) or by any combination of two or more
 telescopes (stereo, CT5 plus at least one of CT1--4, or at least two
of CT1--4). The field of view of CT5 is $3.2\degr$ in diameter,
smaller than that for CT1--4.  Consequently, not all stereo triggers include
CT5. The standard observation mode of \phasetwo is to collect both mono and
stereo events during the same observation run.

The analysis of CT1--5 stereo data provides a lower energy threshold, 
better hadron rejection and better angular resolution than with CT1--4 
only. The analysis of \phasetwo mono events potentially provides a 
factor of $\approx 4$ lower energy threshold than CT1--5 stereo. 
However, the absence of stereoscopic constraints makes the rejection of 
hadronic events more difficult, leading to a larger background and 
reduced signal-to-background ratio at the analysis level. The low 
energy threshold of \phasetwo mono implies high event rates, and thus 
small statistical uncertainties on the background, which leads to tight 
requirements for the accuracy of background subtraction. The 
angular reconstruction of the monoscopic analysis is significantly less 
precise than that obtained in the stereoscopic mode, leading to a 
reduction of the  sensitivity for point-like sources.

Nevertheless, the \phasetwo mono analysis provides new opportunities to 
probe astronomy at energies $<100$~GeV for southern sources, which are 
complementary to satellite experiments (e.g. {\it Fermi} Large Area 
Telescope, LAT) and to northern hemisphere facilities such as MAGIC and 
VERITAS which can detect northern sources below 100~GeV 
\citep{2015JHEAp...5...30A,2015ApJ...815L..22A}. The low energy 
threshold provided by \phasetwo mono is, consequently, particularly beneficial for 
studies of bright variable objects such as gamma-ray bursts and active 
galactic nuclei (AGNs) out to high redshifts ($z \gtrsim 0.5$), 
along with associated spectral features  introduced into the 
spectra through gamma-ray interactions with the extragalactic 
background light (EBL).

The full performance characterization of the CT1--5 system will be
provided in a forthcoming publication.

\section{\phasetwo Mono Observations and Analysis}

\subsection{\phasetwo Observations}
\label{hess2_ct5}

\pks was monitored with \phasetwo regularly for two
consecutive years: in 2013
(from Apr 21 to Nov 5, 2013, MJD 56403--56601);
and 2014 (May 28 -- Jun 9, 2014, MJD 56805--56817).  
\pg was observed with \phasetwo between May 29 and Aug 9, 2013 (MJD 56441--56513).
Most of the observations were taken using the full \phasetwo array.
This paper only reports on the monoscopic analysis of this data, which provides the lowest achievable energy threshold.

\hess data taking is organised in 28 min blocks, called runs.
Observations are usually taken in {\it wobble} mode,
with the camera's field of view centred at a 0.5$\degr$ or 0.7$\degr$ offset
from the source position, in either direction along the right-ascension or declination axis.
Only runs for which the source position is located between 0.35$\degr$ and 1.2$\degr$ off-axis from camera centre are used in the present analysis\footnote{
 Runs with non-standard wobble offests were taken during the commissioning phase to assess the performance of the instrument.}.
This is to ensure that the source is well within the field of view and allow background subtraction using 
the reflected-region background method \citepads{2007A&A...466.1219B}.

\subsection{Data Quality Selection}
\label{data_quality}

To ensure the quality of the AGN data sets for the \phasetwo mono analysis 
the following run quality criteria were applied.
\begin{itemize}
\setlength\itemsep{0em}
\item Stable clear sky conditions according to the telescope radiometers\footnote{We use
the narrow field-of-view radiometers installed on the CT1--4 telescopes,
requesting radiometer temperature to be less than $-$20$\ \degr$C
and stable during the run within $\pm$ 3$\ \degr$C.
};
\item Relative humidity $< 90\%$;
\item Run duration\footnote{A run may be interrupted due to an automated target-of-opportunity observation of a transient source, deteriorating weather conditions, or a technical issue.} $> 5$ min and live time fraction $> 90\%$;
\item At least 90\% of pixels in CT5 are active (pixels can be temporarily switched off due to a star in the field of view or removed from the data due to bad calibration);
\item CT5 trigger in standard configuration pixel/sector threshold = 4/2.5, see \citetads{2006A&A...457..899A} for a definition of the trigger pattern;
\item CT5 trigger rate between 1200 and 3000~Hz (its nominal value depends on the observed field of view and zenith angle)
and stable within $\pm10$\% during a run;
\item Telescope tracking functioning normally;
\end{itemize}

\subsection{Data Analysis}
\label{hess2_analysis}

The data sets were processed with the standard \hess analysis software using the Model reconstruction \citepads{2009APh....32..231D}
which was recently adapted to work with monoscopic events \citepads{2015arXiv150902896H}.
The Model reconstruction performs a likelihood fit of the air shower image to
a semi-analytical model of an average gamma-ray shower parameterised as a function of energy, primary interaction depth, impact distance and direction.
Gamma-like candidate events are selected based on the value of the goodness-of-fit variable
and the reconstructed primary interaction depth.
In addition, events with an estimated error in direction reconstruction $> 0.3\degr$ are rejected.
The low energy threshold is controlled with a dedicated variable \textit{NSB Goodness}, 
which characterises the likelihood of accidentally triggering on fluctuations due to the night sky background.
Two cut configurations were defined for this analysis, {\it loose} and {\it standard}, with different settings for the \textit{NSB Goodness} cut.
{\it Loose} cuts provide the lowest energy threshold, 
 but may lead to a significant level of systematic errors in the background subtraction when applied to high statistics datasets.
{\it Standard} cuts provide a better control over the background subtraction at the cost of increased threshold.
The event selection cuts, except for the \textit{NSB Goodness} cut, were optimised to maximise
the discovery potential for a point source with a photon index of 3.0
observed at a zenith angle of 18$\degr$ for 5~hr.
The optimized analysis provides an angular resolution of $\approx$ 0.15$\degr$ (68\% containment radius) at 100 GeV and energy resolution of $\approx$ 25\%.
 For photon indices harder than 3.0, \textit{standard} cuts provide a better sensitivity than \textit{loose} cuts.

The background subtraction is performed using the standard algorithms used in \hess
 -- the ring background method (for sky maps) and the reflected-region background method 
 \citepads[with multiple off-source regions, for spectral measurements]{2007A&A...466.1219B}.
The ring background method uses a zenith-dependent two-dimensional acceptance model,
an inner ring radius of 0.3$\degr$ and outer radius of 0.6$\degr$, 
and top-hat smoothing radius of 0.1$\degr$.
The acceptance model, which describes the observed distribution of background events in the camera's field of view in absence of gamma-ray sources,
is obtained from the data itself, using background events outside of a radius of 0.3$\degr$ from any known VHE gamma-ray source
(for this analysis, \pks and \pg).
The reflected-region background method uses an on-source region radius of 0.122$\degr$,
which corresponds to an angular distance cut $\theta^2 < 0.015$ deg$^2$.
The number of off-source regions was adjusted on a run-by-run basis
so as to always use the maximum possible number of them, given the wobble angle.
For instance, for a wobble angle of 0.5$\degr$ nine off-source regions were used.
A simple acceptance model, which only corrects for linear gradients in the acceptance, is used with this method.
The significance of the excess after background subtraction is determined using the method described by \citetads{1983ApJ...272..317L}.
Spectral measurements are obtained using the forward folding technique \citepads{2001A&A...374..895P},
applied to the excess events observed with the reflected-region background method.
The energy threshold for the spectral fit is defined as the
energy at which the effective area reaches 15\% of its maximum value, in line with 
the definition previously adopted in \hess analysis \citepads{2014A&A...564A...9H}. 
Such a definition ensures that the systematic uncertainties in the analysis are kept under control.
The \phasetwo mono analysis was applied to all events that include CT5 data (ignoring information from CT1--4).

\section{Results}
\label{section:results}

\subsection{PKS\,2155$-$304}
\label{pks2155}

\begin{figure}
  \centering
  \includegraphics[width=0.49\linewidth]{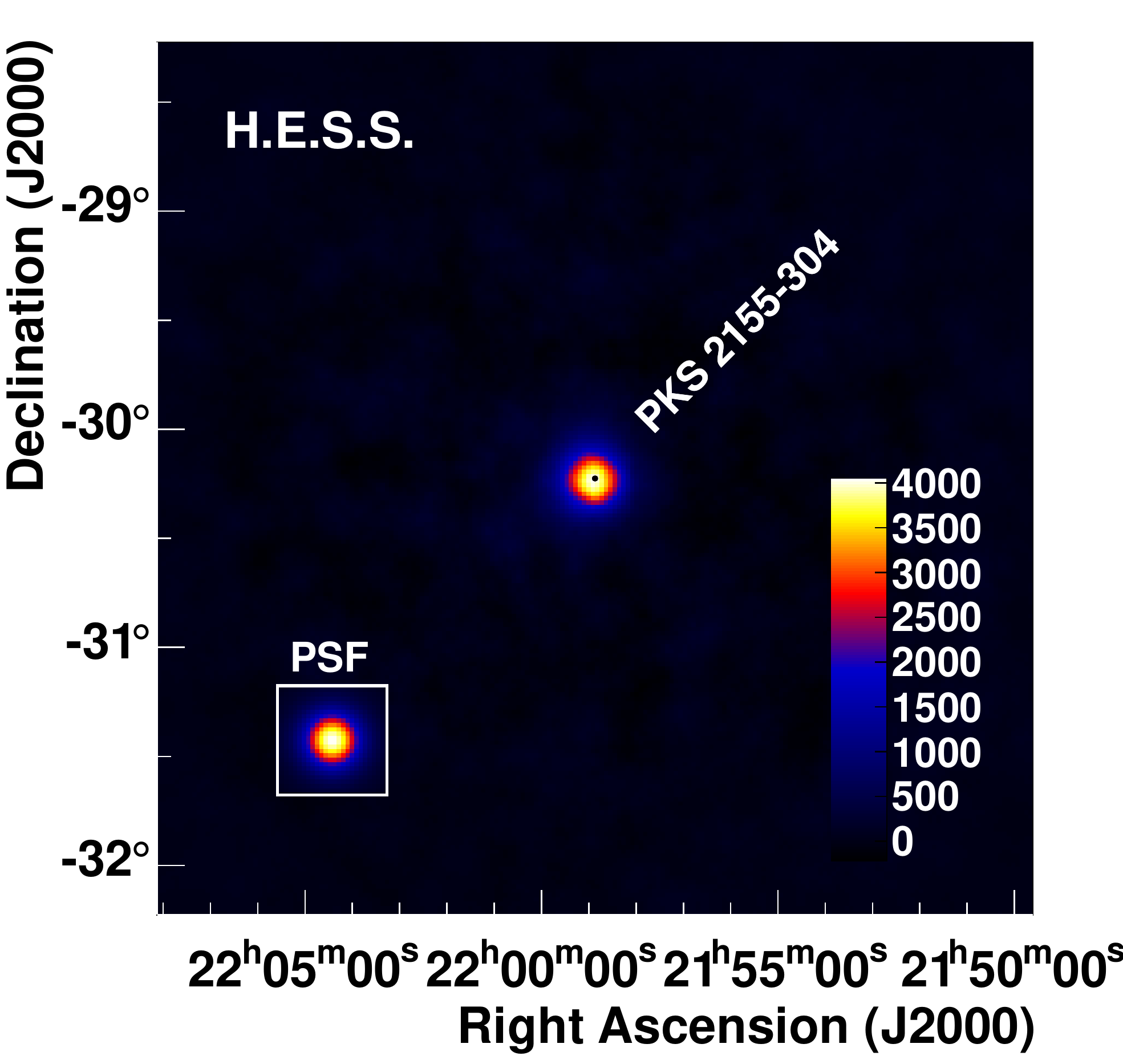}
  \includegraphics[width=0.49\linewidth]{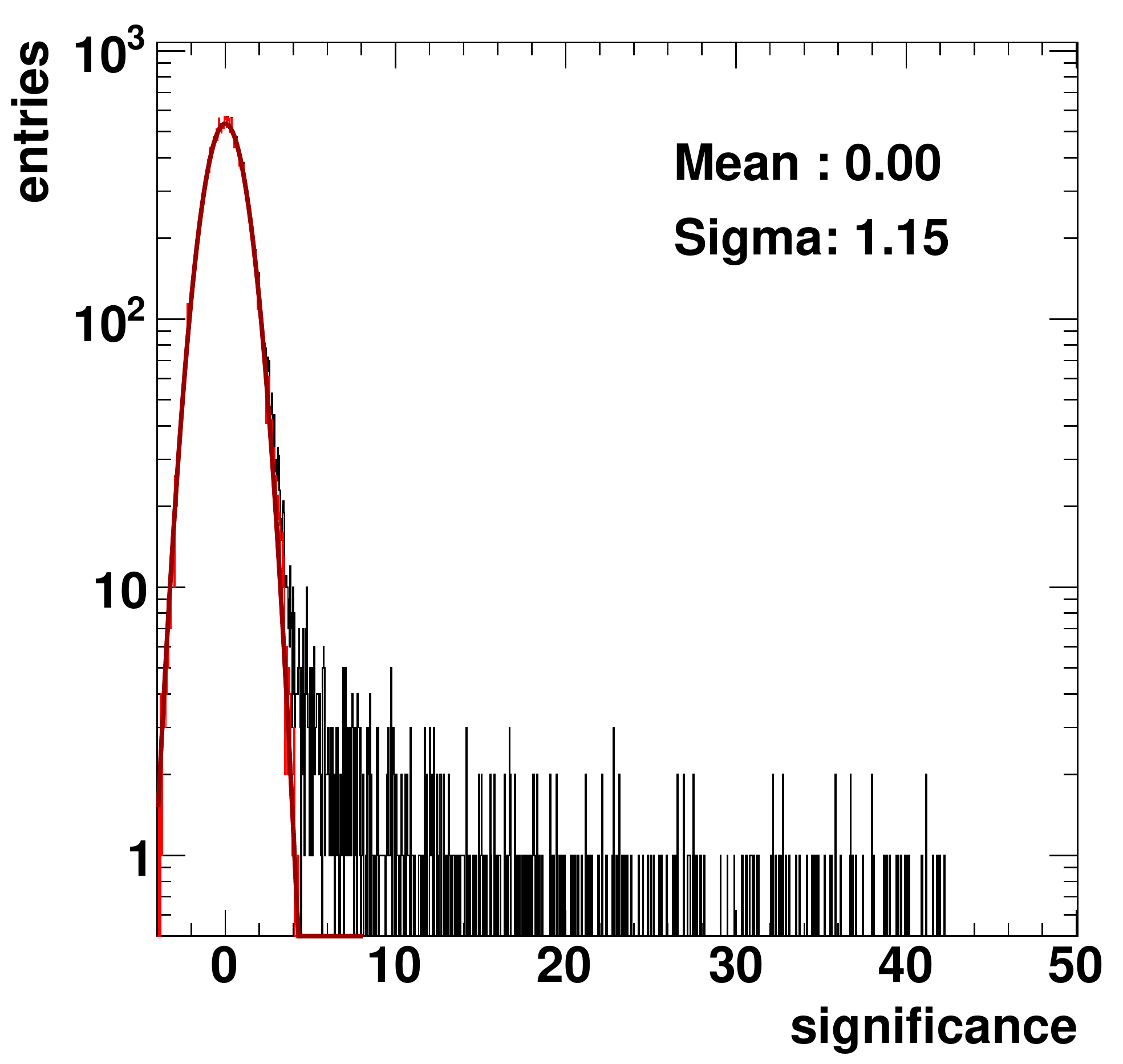}
  \includegraphics[width=0.7\linewidth]{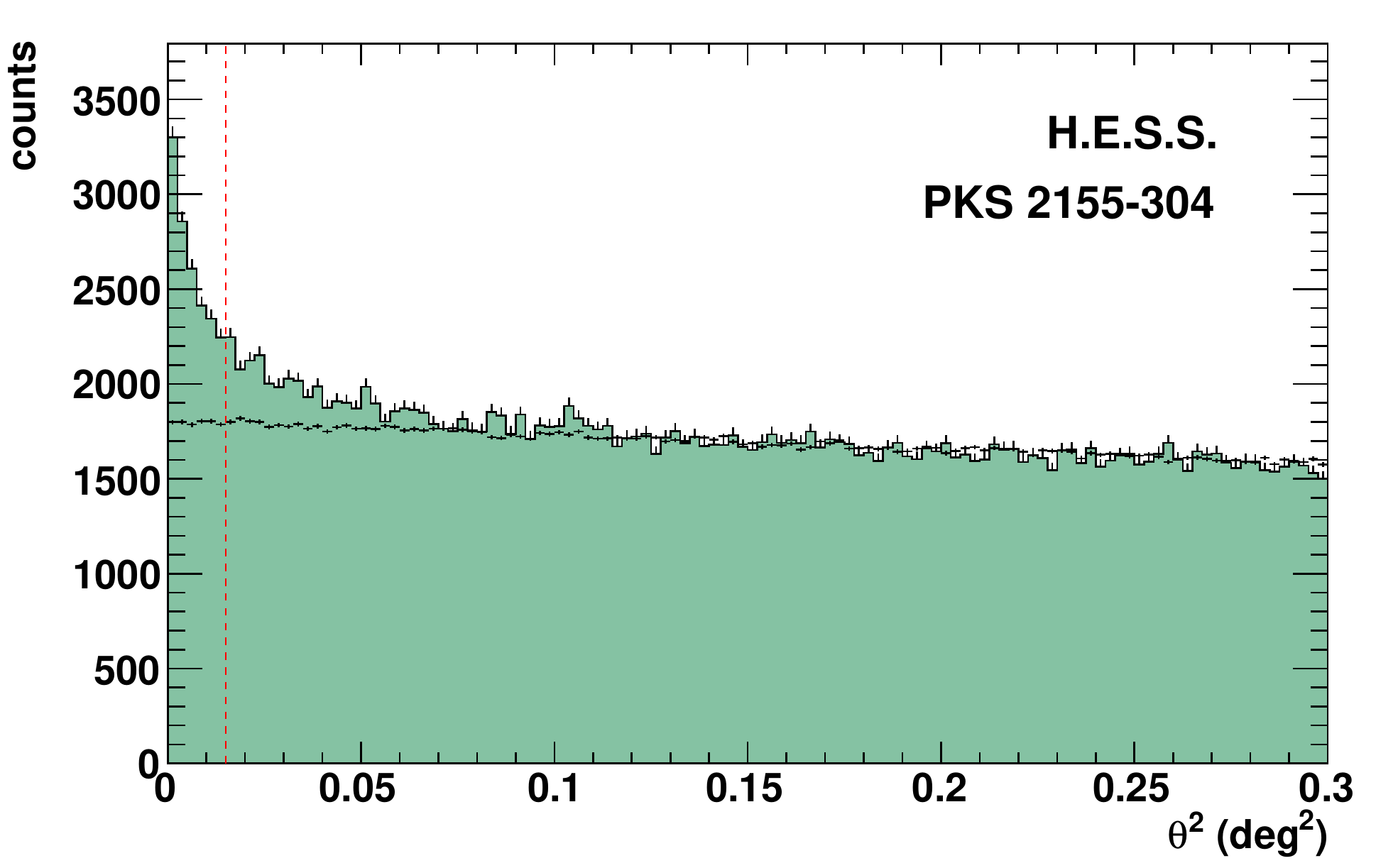}
  \caption{Top: (Left) Excess map of events observed in the direction of \pks
using the \phasetwo mono analysis (2013--2014 data). The inset represents the point spread function of the instrument obtained from simulations.
The source position is indicated by a black dot. (Right) The significance distribution that corresponds to the excess map (black histogram).
The distribution obtained by excluding a circular region of $0.3\degr$ radius around the source is shown in red;
the results of a Gaussian fit to this distribution are also shown. Bottom: The distribution of $\theta^2$ (squared angular distance to \pks) for gamma-like events obtained with the \phasetwo mono analysis (filled histogram)
in comparison with the normalised $\theta^2$ distribution for off-source regions (black points).
The vertical dashed line shows the limit of the on-source region. The energy threshold for this analysis is $\approx$ 80 GeV
}
\label{pks2155_theta2}
\end{figure}

The \pks data set, filtered as explained in Sect.~\ref{data_quality}, comprises 
138 runs. The total live time of this data set is 56.0~hr, 43.7~hr 
taken in 2013 and 12.3~hr taken in 2014.
During these observations, the source zenith angle ranged from 7$\degr$ to 60$\degr$, 
with a median value of $16\degr$.
This data set was analysed using {\it standard} cuts as described in Sect.~\ref{hess2_analysis}.
The background event counts obtained for the off-source regions in each run (in the reflected-region 
background analysis) were used to perform an additional test of the uniformity of the 
camera acceptance. This was done using a likelihood ratio test (LLRT), with the baseline hypothesis that 
the event counts observed in all off-source regions come from the same Poisson distribution, and a nested model allowing for different mean values in each region. The results of this test were consistent with an axially-symmetric 
camera acceptance.

\begin{figure}
  \centering
  \includegraphics[width=0.49\linewidth]{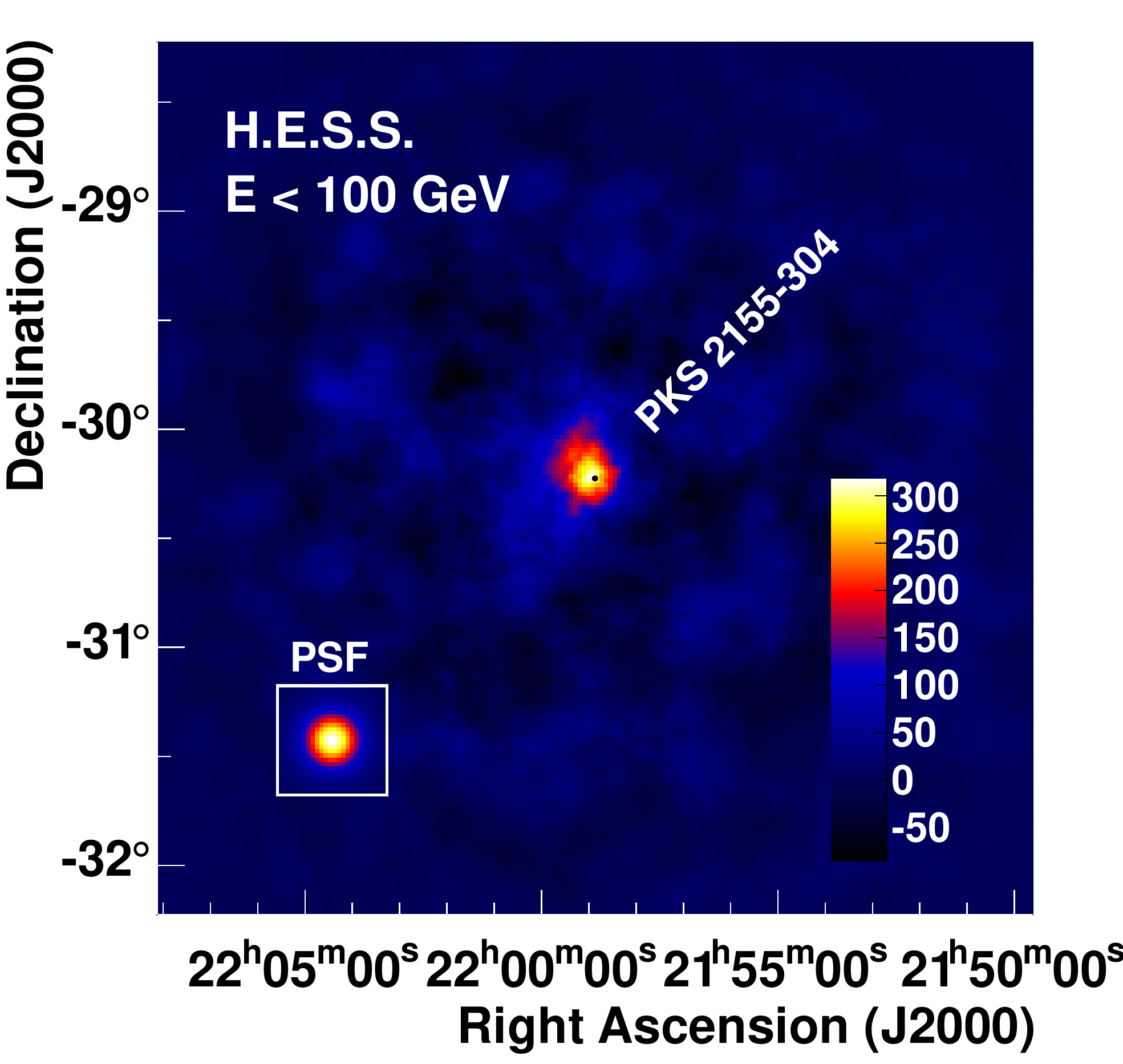}
  \includegraphics[width=0.49\linewidth]{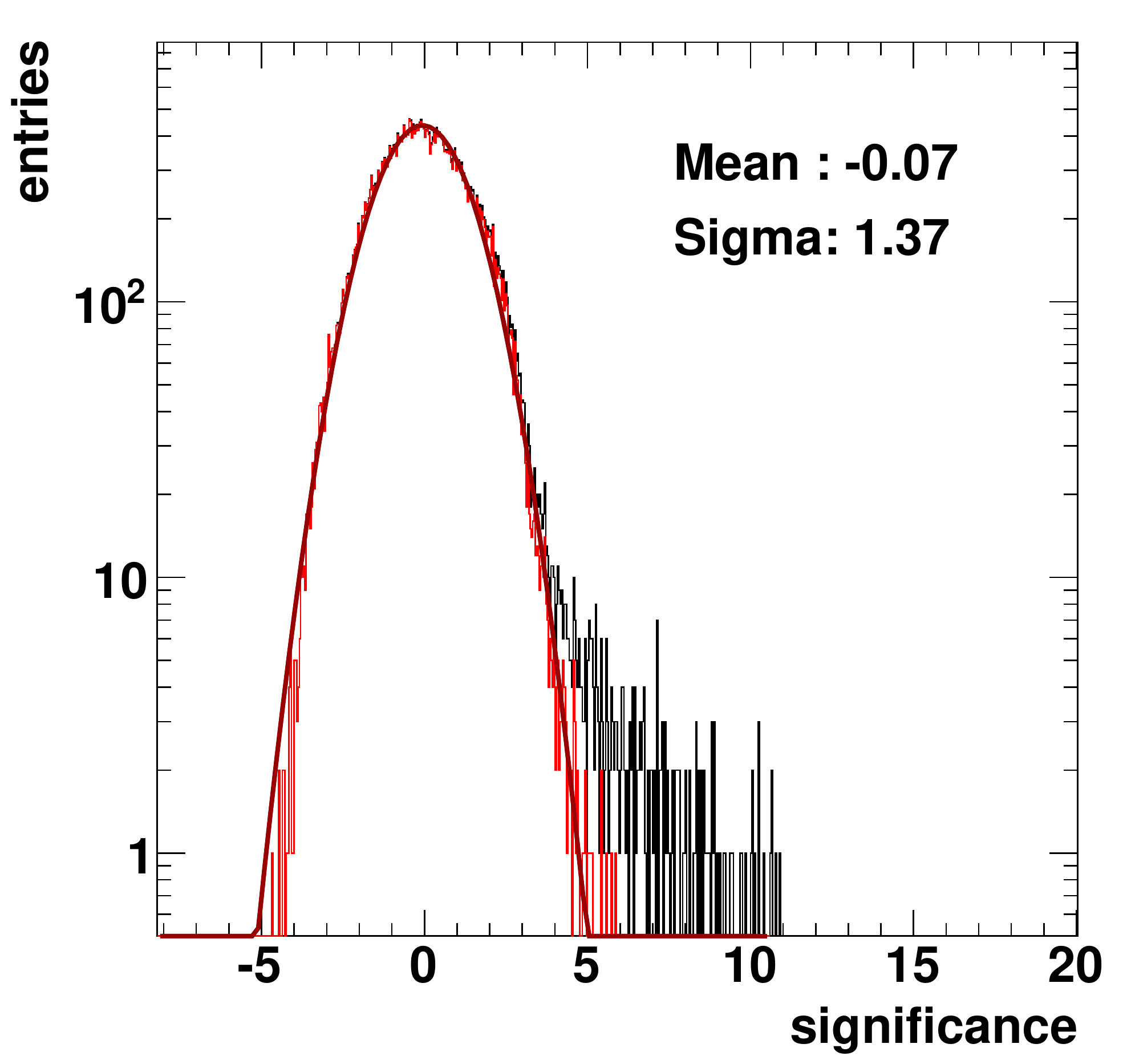}
  \includegraphics[width=0.7\linewidth]{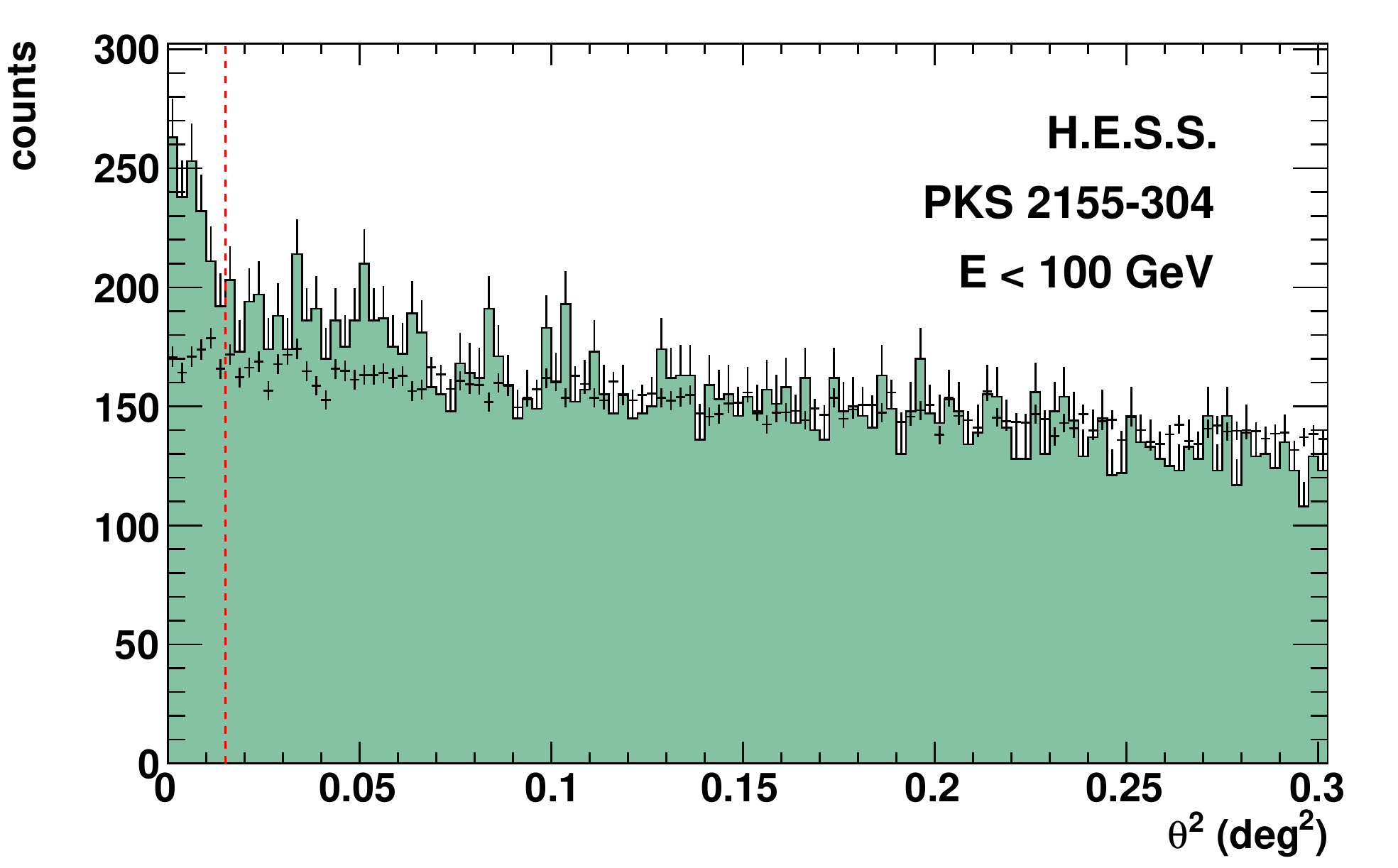}
  \caption{Top : The \pks excess map (left) and significance distribution (right) for events with reconstructed energy $E<100$~GeV
(\phasetwo mono analysis, 2013--2014 data) Bottom: The distribution of $\theta^2$ (squared angular distance to \pks) for gamma-like events.
}
  \label{pks2155_ebins}
\end{figure}

The sky map obtained for \pks using the \phasetwo mono analysis is shown in the top-left panel of 
Fig.~\ref{pks2155_theta2}. The analysis found that the source is detected with a 
significance of $\approx 42~\sigma$, with $\approx$ 4000 excess events. 
The corresponding distribution of the excess significance 
of all skymap bins is shown in the top-right panel of Fig.~\ref{pks2155_theta2}. 
The width of the observed excess is approximately compatible with the simulated point spread function
(PSF; shown in the inset on Fig.~\ref{pks2155_theta2}).
The best-fit position of the excess is found $32\arcsec \pm 10\arcsec_{stat}$ from the target position.
This offset can be attributed to the systematic errors on the telescope pointing.
Outside the exclusion radius 
of $0.3\degr$ the significance distribution was found to be well fit by a Gaussian with 
$\sigma = 1.149 \pm 0.004$. This result indicates the presence of a systematic effect in background 
subtraction, whose $\sigma_{\rm{syst}}$ corresponds to about 57\% of the statistical errors\footnote{
We here assume that the errors add in quadrature.
A value of $\sigma = \sqrt{1 + \sigma_{syst}^2} > \sqrt{2}$
would then indicate the dominance of background subtraction errors.} 
($\sigma_{\rm{stat}} = 1.0$ by construction). 
This effectively reduces the observed excess significance from $42~\sigma$ to $\approx 36~\sigma$\footnote{From this point forward, significance values are not corrected for this effect, with the corrected values being quoted within brackets immediately proceeding these uncorrected values.}.
This systematic effect is currently under investigation as part of a 
larger effort to understand the mono analysis performance. Repeating the analysis using 
only events with reconstructed energy below 100~GeV leads to a $10~\sigma$ ($7.3~\sigma$) significance at 
the position of \pks in the skymap (Fig.~\ref{pks2155_ebins}). The significance 
distribution outside the exclusion region has $\sigma = 1.374 \pm 0.005$,
indicating that the background subtraction errors 
are slightly smaller than the statistical errors.
Thus the source is confidently detected at $E < 100$ GeV.

The distribution of $\theta^2$, the square of the angular difference between the reconstructed 
shower position and the source position, is shown in the bottom panel of Fig.~\ref{pks2155_theta2}
(filled histogram). A $43~\sigma$ excess over the background (black crosses) is observed 
within the on-source region ($\theta^2 < 0.015$ deg$^2$). 

\begin{figure}
  \centering
\includegraphics[width=1.0\linewidth]{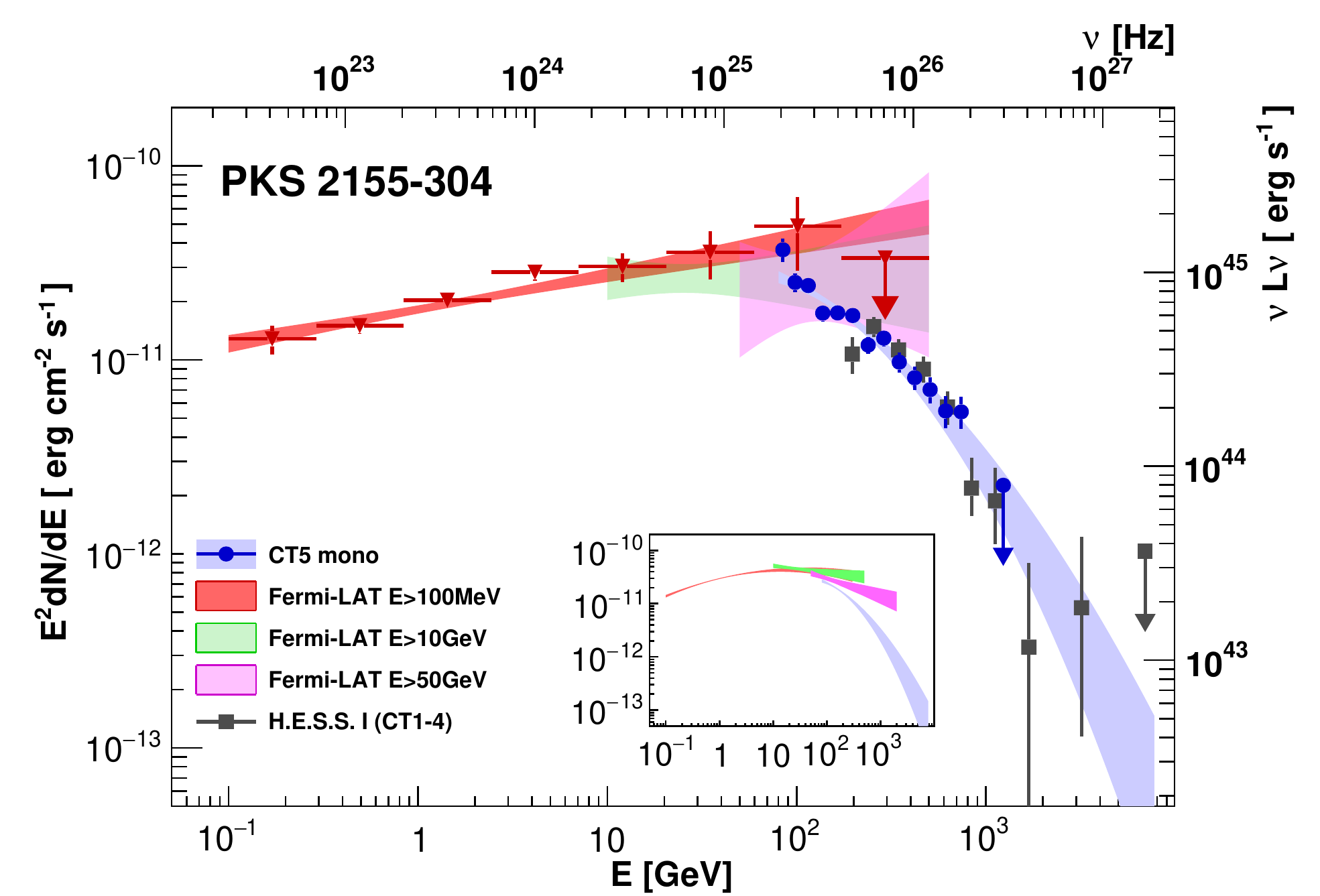}
  \caption{The energy spectrum of \pks obtained from the \phasetwo mono analysis (2013 data, shown by blue circles with confidence band) in comparison with the contemporaneous \fermi data with an energy threshold of 0.1~GeV (red triangles and confidence band), 10~GeV (green band), and 50~GeV (purple band) and contemporaneous CT1--4 data  (grey squares). 
In all cases the confidence bands represent the 1~$\sigma$ region.
The right-hand y-axis shows the equivalent isotropic luminosity (not corrected for beaming or EBL absorption).
The inset compares the \hess confidence band with the \fermi catalogue data (3FGL, 1FHL and 2FHL, see Sect.~\ref{fermi_cat}).}
  \label{pks2155_spectrum}
\end{figure}

\begin{figure}
  \centering
    \includegraphics[width=1.0\linewidth]{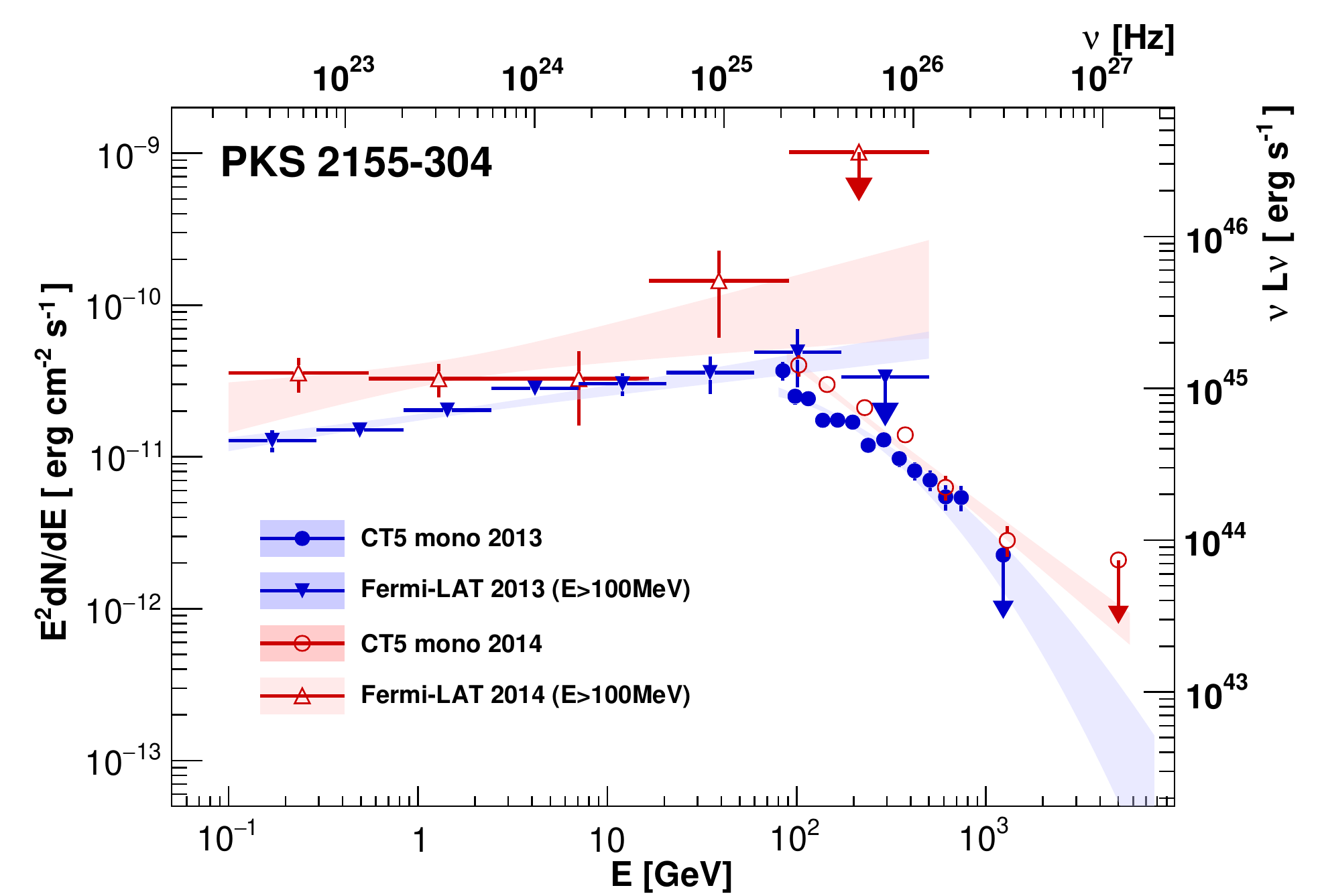}
  \caption{The SED of \pks separated into the 2013 and 2014 observation periods. Both the \phasetwo mono and contemporaneous \fermi data are shown. The bands represent the 1~$\sigma$ confidence region.}
  \label{pks2155_SED_2013_2014}
\end{figure}

The reconstructed spectrum of \pks obtained for 2013, and each of the observation years
(2013 and 2014), is shown in Figs.~\ref{pks2155_spectrum}~and~\ref{pks2155_SED_2013_2014}, respectively. For the full data set (2013+2014), 
a log-parabola model, ${\rm d}N/{\rm d}E = \Phi_{0}\,(E/E_0)^{-\Gamma-\beta\cdot \log(E/E_0)}$,
better fits the data with respect to a simple power-law model with a log-likelihood ratio of 25 (i.e. $5~\sigma$).
The flux normalisation is found to be $\Phi_{0} = (5.11 \pm 0.15_{\rm stat}) \times 10^{-10} \, \rm{cm}^{-2} \, \rm{s}^{-1} \, \rm{TeV}^{-1}$
at a decorrelation\footnote{For the log-parabola model, the decorrelation energy is the energy where the error on the flux is the smallest, i.e. where the confidence band butterfly is the narrowest in the graphical representation.} energy $E_{0} = 156$~GeV,
with a photon index $\Gamma = 2.63 \pm 0.07_{\rm{stat}}$ and a curvature parameter $\beta = 0.24 \pm 0.06_{\rm{stat}}$.
The spectral data points (blue filled circles) cover the energy range from 
80~GeV to 1.2~TeV (not including upper limits).
The spectral parameters obtained for the 2013 and 2014 data sets are given in Table~\ref{TableHESSresults}.
The isotropic luminosity that corresponds to the measured SED is shown by the additional y-axis on the right-hand side of the SED plots.

\begin{table*}[ht]
\caption{Spectral analysis results of \phasetwo mono observations. For both blazars, the observational period is provided along with the spectral parameters: decorrelation energy $E_{0}$; differential flux at the decorrelation energy $\Phi_{0}$; photon index $\Gamma$; and curvature parameter $\beta$. These three parameters describe the log-parabola fit to the spectra.} 
\centering
\begin{tabular}{c c c c c c c c} 
\hline\hline 
Source & Year & MJD & Livetime & $E_{0}$ & $\Phi_{0}$                                                  & $\Gamma$ & $\beta$ \\
       &      &     & [hr] &  [GeV]  & [$10^{-9} \, \rm{cm}^{-2} \, \rm{s}^{-1} \, \rm{TeV}^{-1}$] &          & \\
\hline
\pks & 2013 & 56403--56601 & 43.7 & $151$  & $0.530 \pm 0.018_{\rm stat}$ & $2.65 \pm 0.09_{\rm{stat}}$ & $0.22 \pm 0.07_{\rm{stat}}$ \\
     & 2014 & 56805--56817 & 12.3 & $177$  & $0.532 \pm 0.029_{\rm stat}$ & $2.82 \pm 0.13_{\rm{stat}}$ & $0.16 \pm 0.10_{\rm{stat}}$ \\
     & 2013+2014 & 56403--56817 & 56.0 & $156$  & $0.511 \pm 0.015_{\rm stat}$ & $2.63 \pm 0.07_{\rm{stat}}$ & $0.24 \pm 0.06_{\rm{stat}}$ \\
\hline 
\pg  & 2013   & 56441--56513 & 16.8 & $141$  & $1.48 \pm 0.07_{\rm{stat}}$ & $2.95 \pm 0.23_{\rm{stat}}$ & $1.04 \pm 0.31_{\rm{stat}}$ \\
\hline 
\hline 
\end{tabular}

\label{TableHESSresults}
\end{table*}

\subsection{PG\,1553$+$113}
\label{pg1553}

\begin{figure}
  \centering
    \includegraphics[width=0.49\linewidth]{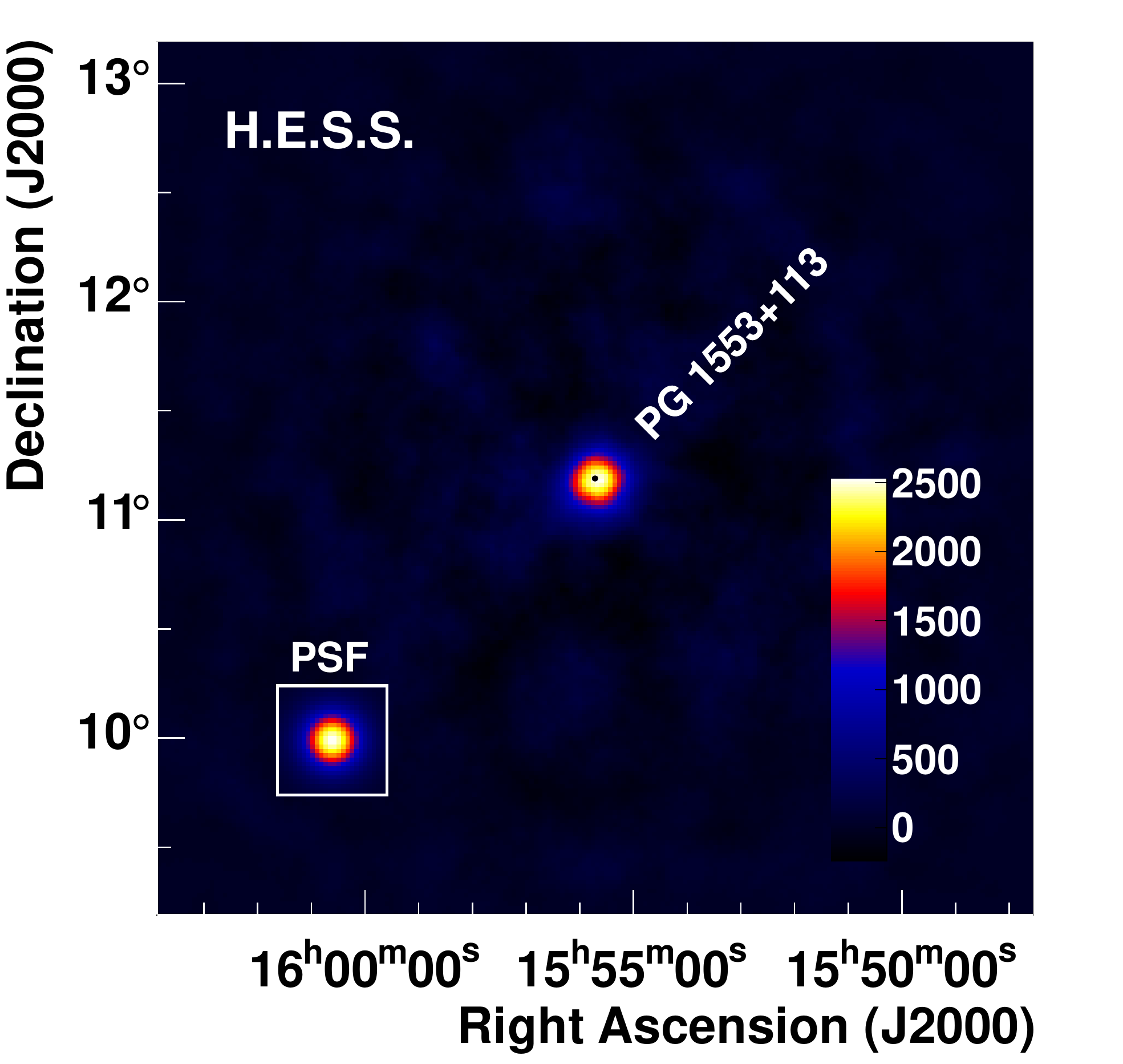}
    \includegraphics[width=0.49\linewidth]{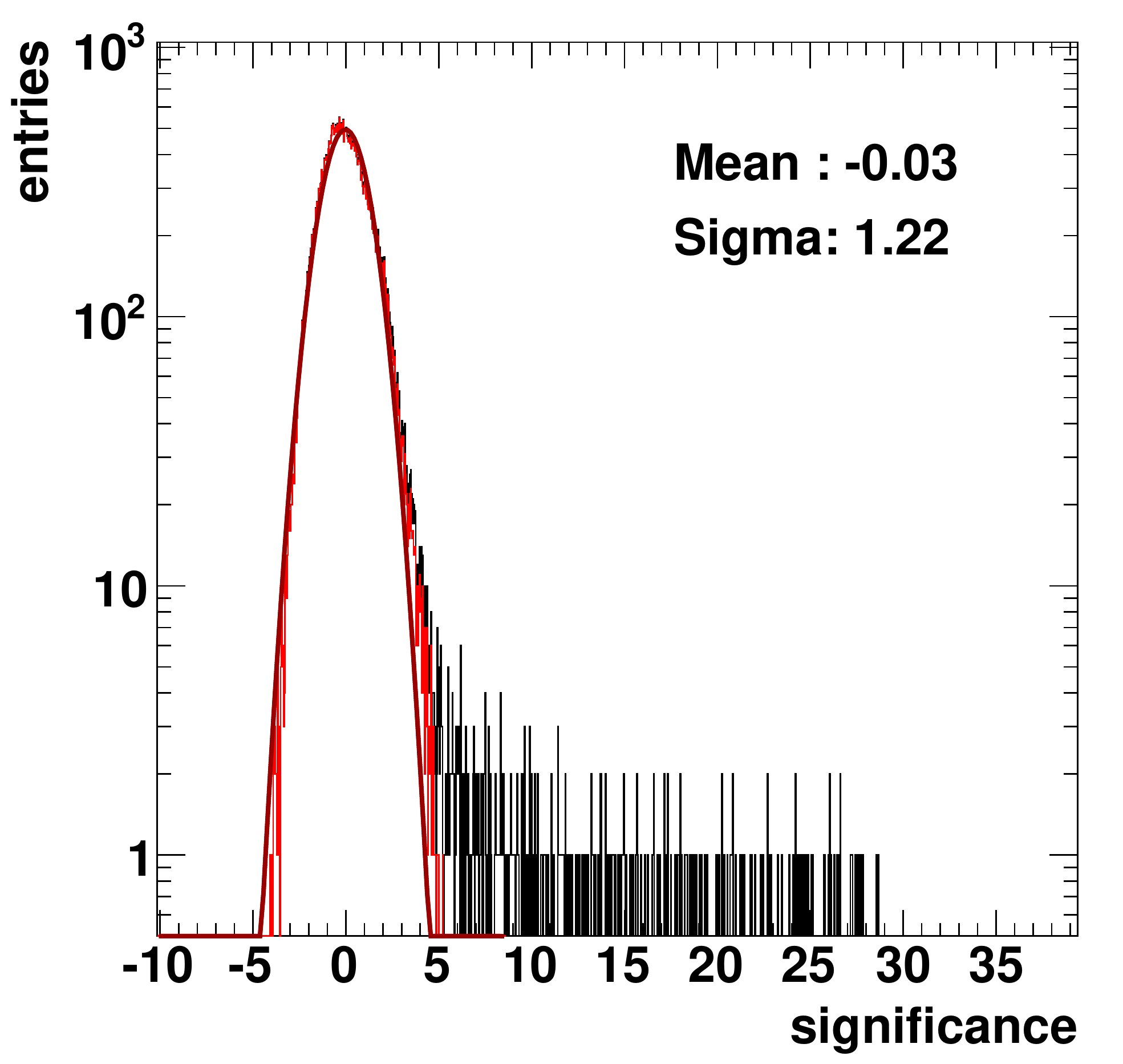}\\
    \includegraphics[width=0.7\linewidth]{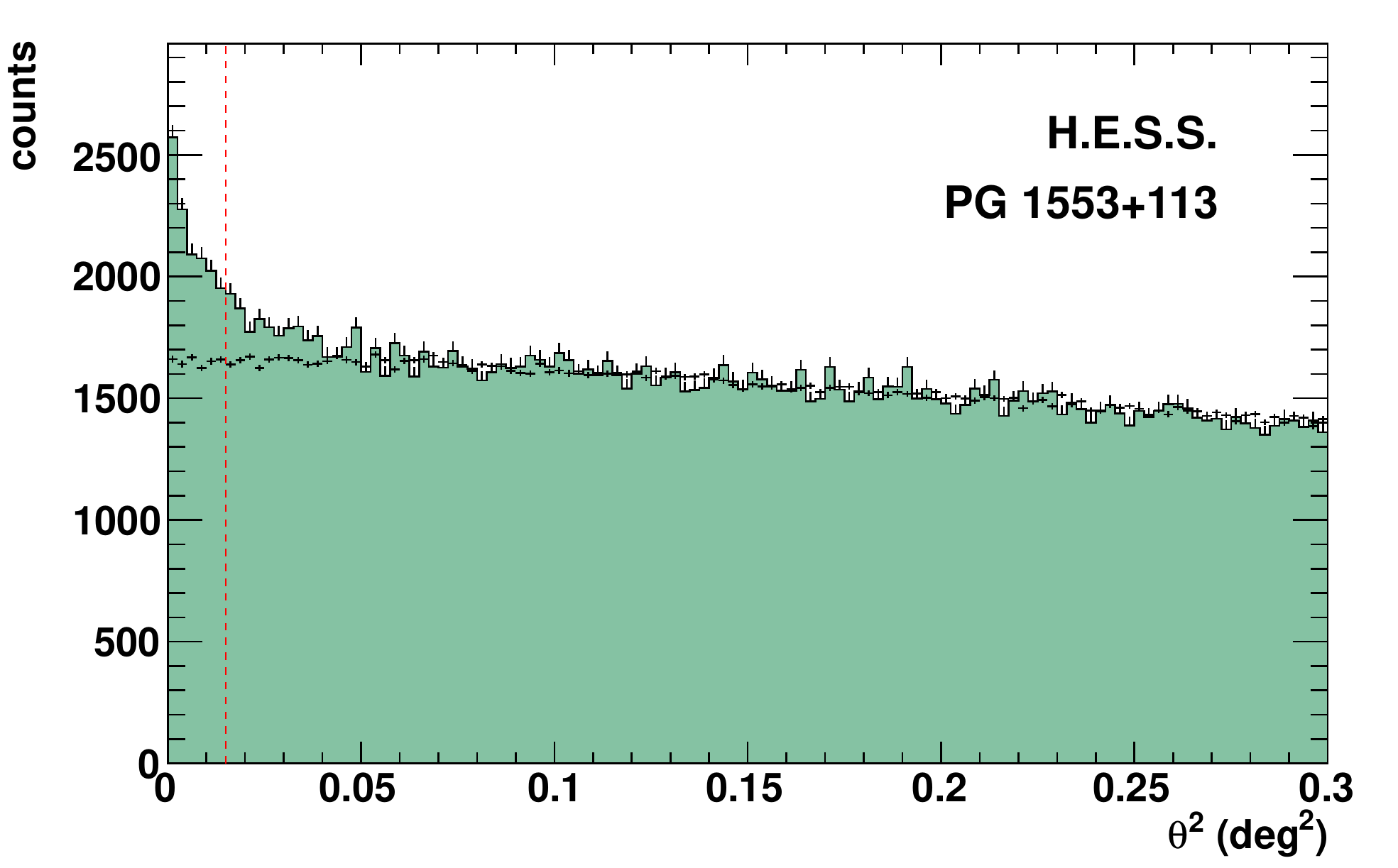}
  \caption{Top: (Left) Excess map of events observed in the direction of \pg
using the \phasetwo mono analysis (16.8 hr live time). The source position is indicated by a black dot. (Right) The significance distribution that corresponds to the excess map.
The meaning of the histograms and statistics data is the same as in Fig.~\protect\ref{pks2155_theta2}. Bottom: The $\theta^2$ distribution for \pg. The meaning of the data shown is the same as in Fig.~\protect\ref{pks2155_theta2}. The vertical dashed line shows the limit of the on-source region.  The energy threshold for this analysis is $\approx$ 100 GeV.
}
\label{pg1553_theta2}
\end{figure}

The \pg data set, filtered as explained in Sect.~\ref{data_quality}, comprises 39 runs (16.8 hr live time), which were analysed using {\it loose} cuts as described in 
Sect.~\ref{hess2_analysis}. This analysis configuration, providing lower energy threshold than {\it standard} cuts, is well suited for bright soft-spectrum sources, such as \pg. During the observations, the source zenith angle ranged 
between $33\degr$ and $40\degr$, with a mean value of $35\degr$. 
The sky map obtained for \pg using the \phasetwo mono analysis is 
shown in the top-left panel of Fig.~\ref{pg1553_theta2}. 
This analysis found that the source is detected with a statistical significance of 27~$\sigma$ (21~$\sigma$), with $\approx$
2500 excess events. 
 
The best-fit position of the excess is found to be $36\arcsec \pm 12\arcsec_{stat}$ from the 
target position, this shift is attributed to the systematic errors on the telescope pointing.
The width of the observed excess is compatible with the simulated PSF within a 10\% systematic uncertainty on the PSF width.

\begin{figure}
  \centering
    \includegraphics[width=0.49\linewidth]{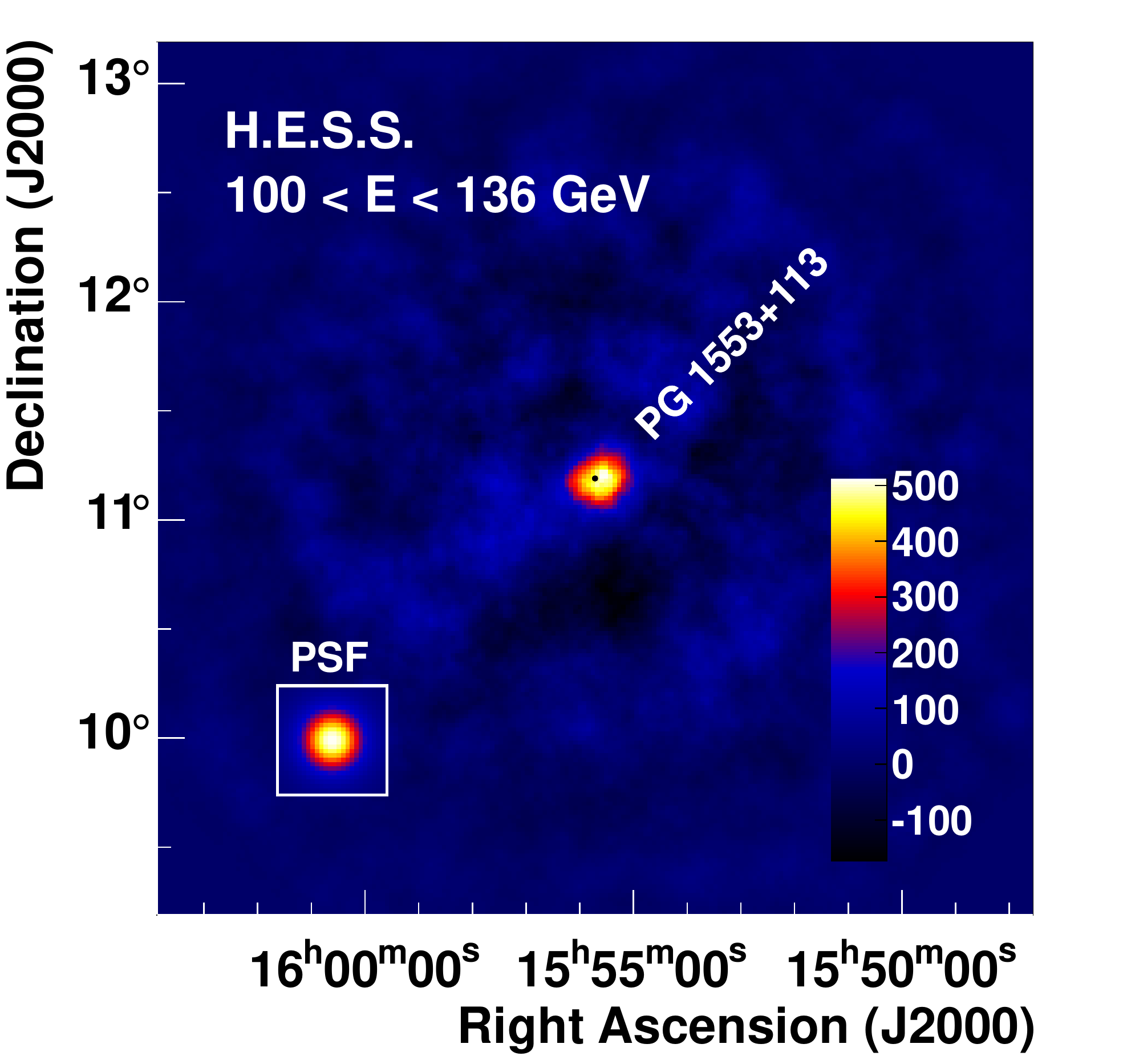}
    \includegraphics[width=0.49\linewidth]{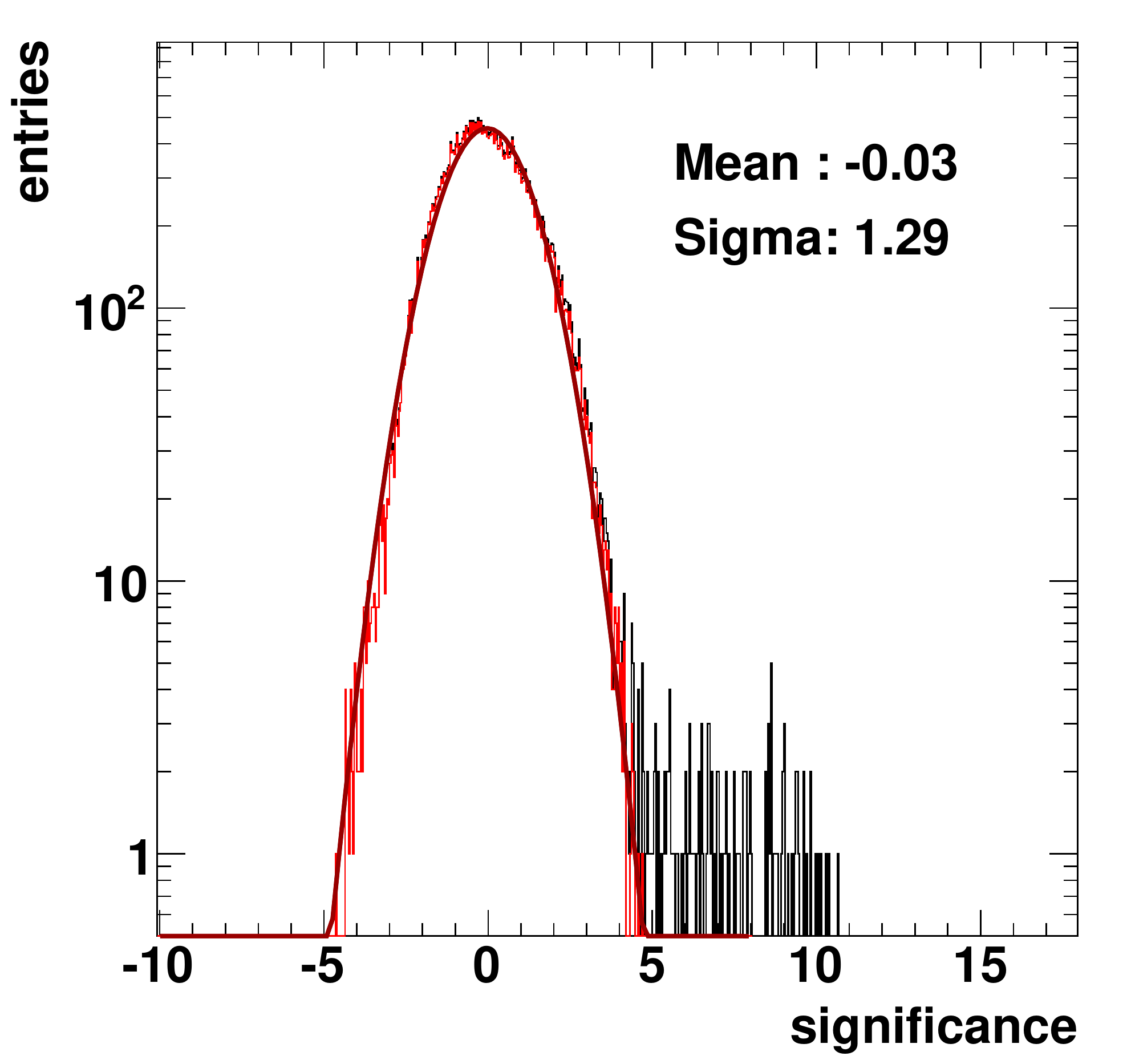}    
   \includegraphics[width=0.7\linewidth]{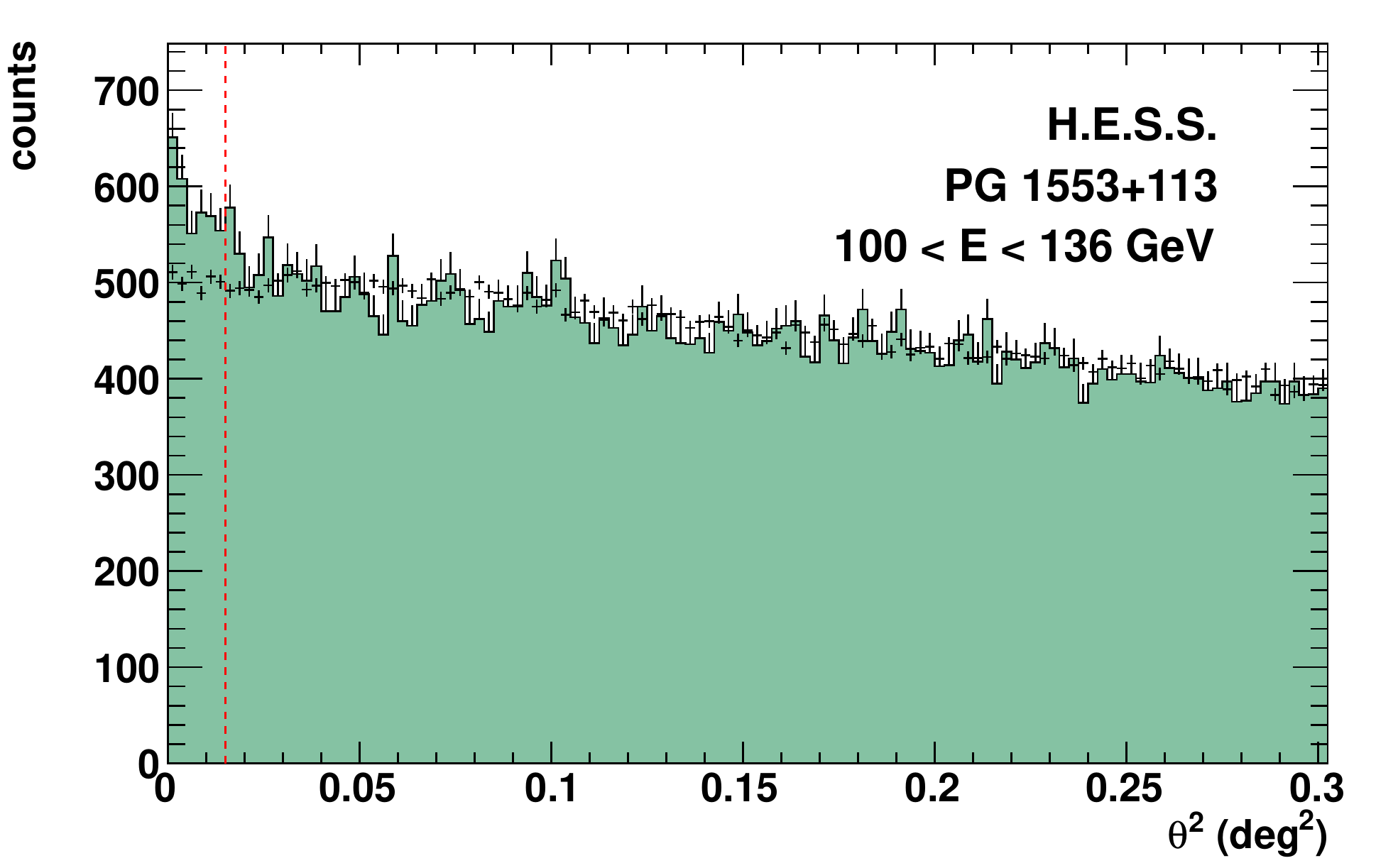}
  \caption{Top : (Left) The \pg excess map and (Right) significance distribution for events with reconstructed energy between 100~GeV and 136~GeV (\phasetwo mono analysis). Bottom: The distribution of $\theta^2$ (squared angular distance to \pks) for gamma-like events.
}
  \label{pg1553_ebins}
\end{figure}

The significance distribution in the region outside of the $0.3^\circ$ exclusion 
radius is consistent with a normal distribution (top-right panel of Fig.~\ref{pg1553_theta2}).
The same holds true when the analysis is repeated in only a low energy bin,
with a reconstructed energy range of 100 -- 136~GeV. Within this energy bin, the source is 
detected with a 10~$\sigma$ (8.2~$\sigma$) significance (Fig.~\ref{pg1553_ebins}). 
The significance distribution outside the 
exclusion region has $\sigma = 1.219 \pm 0.005$ and $1.288 \pm 0.005$,
for the full energy range and the first energy bin, respectively,
indicating presence of background subtraction errors 
at a level smaller than the statistical errors.

The $\theta^2$ distribution is shown in the bottom panel of Fig.~\ref{pg1553_theta2}.
A 27~$\sigma$ (21~$\sigma$) excess over the background is observed within the on-source region 
($\theta^2 < 0.015$ deg$^2$). The reconstructed spectrum, with a threshold of 110~GeV, is found to be well fit by a log-parabola
(with a LLRT of 20 over the power-law model, Fig.~\ref{pg1553_spectrum}), with a photon index $\Gamma = 2.95 \pm 0.23_{\rm{stat}}$ 
at decorrelation energy $E_{0} = 141$~GeV, curvature parameter $\beta = 1.04 \pm 0.31_{\rm{stat}}$,
and differential flux $\Phi_{0} = (1.48 \pm 0.07_{\rm{stat}}) \times 10^{-9} \, \rm{cm}^{-2} \, \rm{s}^{-1} \, \rm{TeV}^{-1}$ at $E_{0}$.
The spectral data points (blue filled circles) cover the energy range from 110~GeV to 
550~GeV	 (not including upper limits).

\subsection{Cross check analysis}\label{sect:crosscheck}
The robustness of the new \phasetwo mono results presented above has been tested through an 
independent analysis using the Image Pixel-wise fit for Atmospheric Cherenkov Telescopes 
(ImPACT) method described in \citetads{2014APh....56...26P}. This independent analysis provides a consistent cross-check with the above results, being successfully 
applied to the reconstruction of data coming from CT5-only triggers \citepads{2015arXiv150906322P}.
The analysis was equally capable of detecting \pks below 100~GeV
and the derived spectra were found to be in very good agreement with the Model analysis
for both \pks and \pg. Furthermore, the difference between the spectral parameters derived 
using ImPACT and the Model analysis was adopted as an estimate of the systematic uncertainties 
associated with the reconstruction and analysis techniques (see Sect.~\ref{systematics}).

Additionally, the robustness of the analysis was tested using an alternative 
cut configuration.
Within the statistical and systematic uncertainties, 
the results obtained with the different cut configurations
were found to be in good agreement with each other.

The CT1--4 stereoscopic data collected simultaneously with the  \phasetwo mono data
have been analysed using the \hess~I version of the Model analysis method 
\citep{2009APh....32..231D} using the {\it loose
cuts} \citep{2006A&A...457..899A} to ensure a low energy threshold. In total, data sets of 27.2 hr of live time for \pks and 9.0 hr for \pg have been analysed, yielding a significance of 46~$\sigma$ for \pks and 9.0~$\sigma$ for \pg. 
 Note that the live times differ from the corresponding mono analysis live times due to different run qualities and observation schedules for the different instruments.
For each data set the spectrum is well fitted by a power-law model
and the resulting forward-folded data points for \pks (2013 data) and \pg are shown on Figs.~\ref{pks2155_spectrum}~and~\ref{pg1553_spectrum}, respectively.
The CT1--4 results for \pks were found to be in excellent agreement with the \phasetwo mono results.
Due to the limited statistics and relatively high energy threshold of the CT1--4 analysis,
the CT1--4 results for \pg are represented on Fig.~\ref{pg1553_spectrum} by 3 data points only.
Taking into consideration the systematic uncertainties on the energy scale and flux normalization (see Sect.~\ref{systematics}),
the CT1--4 data were found to be in satisfactory agreement with the CT5 results.

\subsection{HE Gamma-Rays Observed by \fermi}
\label{fermi_data}

\subsubsection{Contemporaneous Data}
\label{fermi_contemp}

\begin{table*}[ht]
\caption{\fermi spectral analysis results for the time intervals contemporaneous with the \phasetwo observations. For each data set and energy threshold, $E_{\rm th}$, the differential flux $\phi_{0}$ at decorrelation energy $E_{0}$, photon index $\Gamma$, and value of the test statistic (TS), for the power-law fit, are provided.} 
\centering
\begin{tabular}{c c c c c c c c} 
\hline\hline 
Source & Year & MJD&  E$_{\rm th}$ &$\phi_{0}$  & $\Gamma$ & E$_{0}$ & TS \\
&  & & (GeV)& $10^{-11}$(ph cm$^{-2}$s$^{-1}$GeV$^{-1}$)  &  & (GeV)&\\
\hline

\pks & 2013 & 56403--56601& 0.1 & 557$\pm$ 26& 1.82$^{+0.03}_{-0.03}$ &1.48&2162.6 \\
 &  & & 10& 2.52 $\pm$0.43&  2.00$^{+0.21}_{-0.21}$ &25.5&379.7\\
 &  & & 50 & 0.12$\pm$0.05&   1.82$^{+0.66}_{-0.72}$ &112&52.4 \\

\pks & 2014 & 56805--56817& 0.1 & 996$\pm$168&  1.79$^{+0.13}_{-0.13}$ &1.54&193.5 \\
 &  & & 10 &  2.36$\pm$1.18 & 1.20$^{+0.45}_{-0.45}$ &53.3&52.4\\
 &  & & 50 & 1.00$\pm$0.71 &1.53$_{-1.20}^{+1.03}$&115&23.7\\
\hline

\pg  & 2013 & 56403--56817& 0.1 & 118$\pm$13& 1.59 $^{+0.07}_{-0.07}$ &  2.95 &455.6\\
 &  & & 10 & 2.04$\pm$0.53& 1.68$_{-0.21}^{+0.26}$&  33.5&169.9 \\
 &  & & 50 & 0.64$\pm$0.27& 2.97$_{-1.13}^{+ 0.91}$&  80.8& 66.8\\
\hline 
\hline 
\end{tabular}
\label{FermiTab}
\end{table*}

The \fermi detects gamma-ray photons above an energy of 100~MeV. Data 
taken contemporaneously with the \phasetwo observations were analysed 
with the publicly available ScienceTools {\tt v10r0p5}\footnote{See 
\url{http://fermi.gsfc.nasa.gov/ssc/data/analysis/documentation/}.}. 
Photon events in a circular region of 15$\degr$ radius centred on the 
position of sources of interest were considered and the {\tt PASS 8} 
instrument response functions (event class 128 and event type 3) 
corresponding to the {\tt P8R2\_SOURCE\_V6} response were used together 
with a zenith angle cut of 90$\degr$. The analysis was performed using 
the {\tt Enrico} Python package \citepads{2013arXiv1307.4534S} adapted 
for {\tt PASS 8} analysis. The sky model was constructed based on the 
3FGL catalogue \citepads{2015ApJS..218...23A}. The Galactic diffuse 
emission has been modeled using the file {\tt gll\_iem\_v06.fits} 
\citepads{2016ApJS..223...26A} and the isotropic background using {\tt 
iso\_P8R2\_SOURCE\_V6\_v06.txt}.

Three energy ranges were considered with the corresponding data cuts in this analysis: 0.1~GeV--500~GeV, 
10~GeV--500~GeV and 50~GeV--500~GeV, with time windows chosen to coincide with 
the \phasetwo observation periods (as defined in Sect.~\ref{hess2_ct5}).

The spectral fit parameter results are given in Table~\ref{FermiTab}. 
For both AGNs a log-parabola fit to the contemporaneous \fermi data
did not provide a sufficient improvement to the spectral fit, 
with respect to the power-law model. Some evidence for a 
softening of the spectrum with energy in the \fermi energy range,
however, was suggested by the analysis of \fermi data for the
scan of energy thresholds shown in 
Figs.~\ref{pks2155_spectrum}~and~\ref{pg1553_spectrum} 
whose fit indices are given in Table~\ref{FermiTab}. The data points have been obtained by redoing the \fermi analysis in a restrained energy range freezing the spectral index of the power-law model to the value found for the global fit above 100~MeV. An upper-limit at 95\% confidence level is computed if the TS is found to be below 9.

These \fermi analysis results are used to provide 
gamma-ray HE-VHE SEDs of \pks and \pg. In Fig.~\ref{pks2155_spectrum}, 
the 2013 \phasetwo data set of \pks is presented along with the contemporaneous 
\fermi data analysed above 100~MeV (shaded red), 10~GeV (shaded green) and 
50~GeV (shaded magenta) respectively. These results show very good agreement 
between the \fermi and \phasetwo mono data within the common overlapping region\footnote{80-500~GeV for \pks and 110-500~GeV for \pg.},
presenting a comprehensively sampled SED over more than four orders of magnitude 
in energy. Evidence for a strong down-turn spectral feature within this
broadband SED, occurring near the transition zone between the two 
instruments, is apparent. 

Figure~\ref{pg1553_spectrum} presents the SED of \pg obtained from
the contemporaneous \fermi and \phasetwo data.
In this case, again, good agreement between the \fermi and \phasetwo mono
data is found within the common energy range of the two instruments.
Furthermore, evidence of a strong down-turn feature within this SED, 
occurring within the overlapping energy range of the two instruments,
is once again apparent.

\subsubsection{Catalogue Data}
\label{fermi_cat}

\begin{figure}
  \centering
\includegraphics[width=1.0\linewidth]{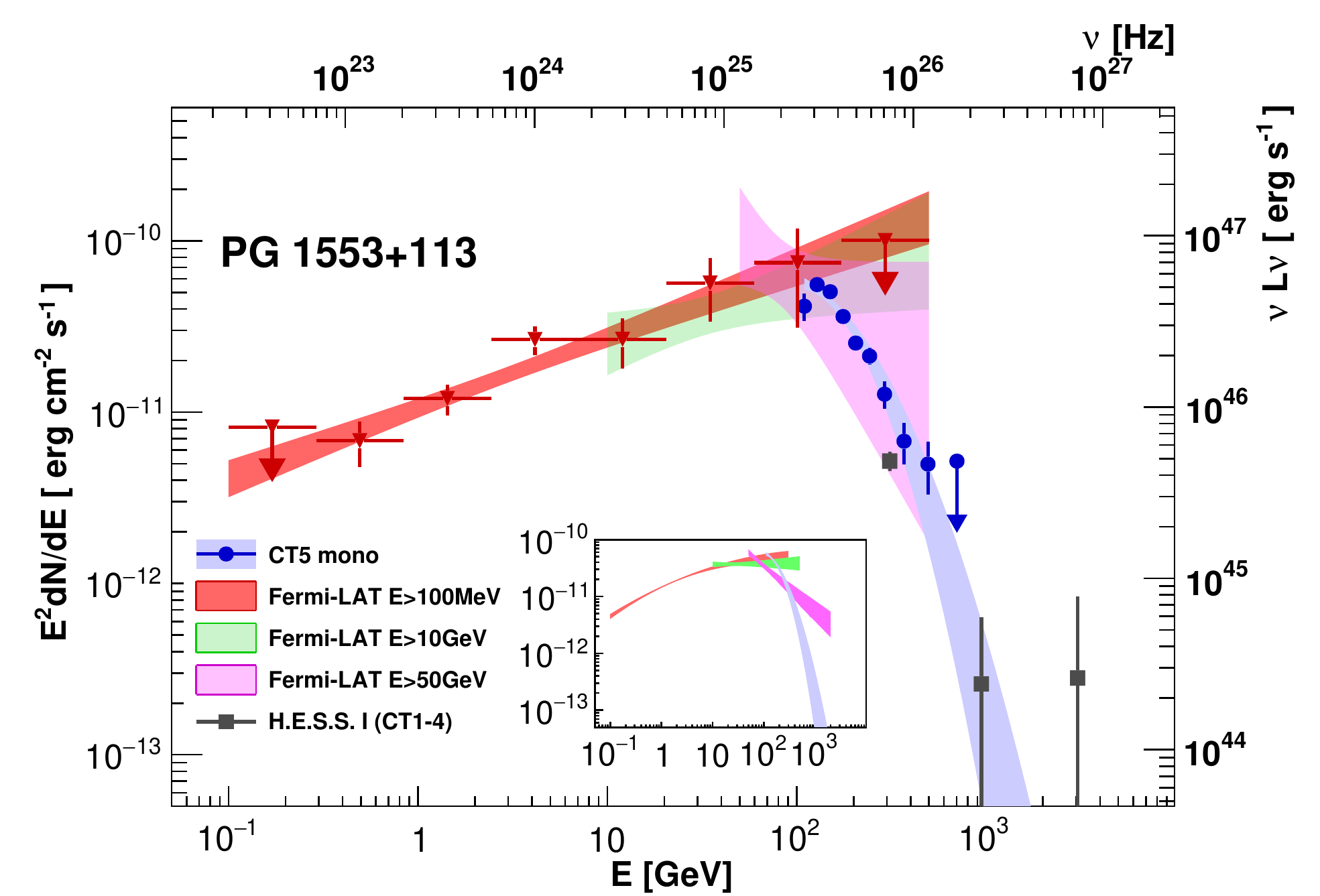}
  \caption{The energy spectrum of \pg obtained from the \phasetwo mono analysis (blue) in comparison with the contemporaneous \fermi data with an energy threshold of 0.1~GeV (red triangles and confidence band), 10~GeV (green band), and 50~GeV (purple band) and contemporaneous CT1--4 data (grey squares).
In all cases the bands shown represent the 1~$\sigma$ confidence region.
The right-hand y-axis shows the equivalent isotropic luminosity (not corrected for beaming or EBL absorption) assuming redshift $z=0.49$.
The inset compares the \hess confidence band with the \fermi catalogue data (3FGL, 1FHL and 2FHL, see Sect.~\ref{fermi_cat}).}
  \label{pg1553_spectrum}
\end{figure}

The \phasetwo mono and contemporaneous \fermi spectra of \pks and \pg obtained in the 
previous sections are compared here to the \fermi catalogue results. Different 
catalogues probing different photon statistics and energy ranges are considered 
here, namely the 3FGL \citepads{2015ApJS..218...23A}, the 1FHL 
\citepads{2013ApJS..209...34A} and the 2FHL \citepads{2016ApJS..222....5A}. The 3FGL 
catalogue gives an average state of the sources with 4 years of data integrated 
in the \fermi analysis above 100~MeV, while the 1FHL relies on the first 3 
years of data with a higher energy cut at 10~GeV. Moreover, the 2FHL 
catalogue was built with the highest energy available to \fermi only, with 
$E>50$~GeV, probing a somewhat different energy range, and thus potentially 
different spectral properties with respect to the FGL source catalogues.

The insets in Figs.~\ref{pks2155_spectrum} and \ref{pg1553_spectrum}
provide a comparison of the \phasetwo mono results (shown by the blue band) with the
\fermi catalogue data (red for 3FGL, green for 1FHL, and purple for 2FHL),
for \pks and \pg, respectively.

It is worth comparing the \fermi contemporaneous 
data obtained in Sect.~\ref{fermi_contemp} and the \fermi catalogue data discussed here.
For \pks, it is noted that the \fermi catalogue flux is slightly above
the \fermi contemporaneous flux in the high energy band.
For \pg, however, the catalogue flux is in close agreement with the \fermi contemporaneous
flux in the high energy band. Since the \fermi 
catalogue data represent the average flux state of the source since data taking 
commenced in 2008, the comparable level of the fluxes (though slightly below for the
case of \pks) is suggestive that 
both sources were in average states of activity during the observational campaign. Note that the catalogues are based on different time intervals and different energy ranges. Furthermore, the results of the fits are dominated by the lower energy events and, in particular for the 2FHL, the statistics are rather poor at the highest energies.

\subsection{Variability}
\label{lightcurves}

The AGNs considered in this work are known to be variable at VHE, both 
having previously been observed to exhibit major flares 
\citepads{2007ApJ...664L..71A,2015ApJ...802...65A}. In the case of \pks, 
this variability has been shown to also introduce changes in the spectral shape 
\citepads{2010A&A...520A..83H}.

In both cases, the present observational campaign found the AGNs to be 
in low states. For \pks, at $E > 300$~GeV the spectrum level from our 
new \phasetwo mono result agrees with the level reported for the 
quiescent state observed by \hess from observations during 2005--2007 
\citepads{2010A&A...520A..83H}. As seen in Fig.~\ref{pks2155_spectrum}, 
at $E < 300$~GeV the \phasetwo mono spectrum level lies below the 
\fermi spectra reported in the 3FGL and 1FHL catalogues. 
These comparisons are all consistent with \pks being in a low flux 
state during the observations analysed in this work, as is also 
indicated by the \fermi contemporaneous analysis results.

Although observed in a low state, the \phasetwo mono lightcurve of \pks 
did exhibit nightly and monthly variability with a fractional variability amplitude $\fvar$ \citepads{2003MNRAS.345.1271V} of, respectively $\approx 47$\% and  $\approx 59$\%. Inter-year variability at VHE with a fractional variability amplitude $\fvar$ of $\approx 50$\% has also been found. Analysis of this variability
in the \phasetwo mono data set revealed that an increase in the flux exists between the 2013 and the 2014 dataset by a factor $1.6\pm 0.1_{\rm{stat}}$, 
though without significant change in the spectral parameters. A simple 
power-law fit to the 2013 (resp. 2014) data yields a spectral index 
$\Gamma_{2013}= 2.92\pm0.04_{\rm{stat}}$ (resp.~$\Gamma_{2014}=2.91\pm0.08_{\rm{stat}}$).
We note, however, that the statistics of the 2013 and 2014 \pks \phasetwo mono data 
sets are significantly different in size. Consequently, the 2014 \pks data set 
is not sufficient to discriminate between a power-law or a log-parabola shaped 
spectrum, whereas the 2013 \pks data set is found to be significantly better fit 
with a log-parabolic spectrum.

For comparison, variability analysis of the \pks contemporaneous \fermi data,
discussed in Sect.~\ref{fermi_contemp}, was carried out.
Fig.~\ref{pks2155_SED_2013_2014} shows the \pks 2013 and 2014 multi-wavelength 
SED obtained. It is notable that a brightening of the source 
flux between these two epochs by about the same level as that seen by 
\phasetwo mono is also observed in the \fermi contemporaneous results, 
and again without any corresponding spectral variability. 
That is the \fermi and \phasetwo mono photon 
indices are respectively consistent between the two epochs, but the overall 
flux increased by about 60\%.

The variability in HE has also been probed on a weekly timescale which gives a good balance between the ability to probe short timescale variations and good statistics. For the 2013 dataset (the 2014 dataset time range being too short), \pks is found to be variable with $\fvar= 37$\%.

For \pg, our new \phasetwo mono spectral results are in reasonable agreement with the earlier 
measurements by \hess \citepads{2008A&A...477..481A,2015ApJ...802...65A} (at $E > 200$~GeV), MAGIC \citepads{2007ApJ...654L.119A,2010A&A...515A..76A,2012ApJ...748...46A} and VERITAS \citepads{2015ApJ...799....7A}, as well as with the \fermi catalogue 
spectra (at $E < 200$~GeV). These comparisons with previous measurements indicate
that \pg was indeed in a low state during the \phasetwo observation period of the results presented. No significant night-by-night
or weekly variability is found in the \phasetwo mono lightcurve. The upper limit on $\fvar$  is found to be 21\% at the 95\% confidence level. In the HE range, \pg is not variable and $\fvar< 110$\% at 95\% CL.

\section{Systematic Uncertainties}
\label{systematics}

\begin{table*}[ht]
\caption{Estimated contributions to the systematic uncertainties in the spectral measurements using \phasetwo mono for the analyses presented in this work.
Numbers separated by "/" correspond to \pks and \pg, respectively.}
\label{SystErrorTable}
\centering
\begin{tabular}{c c c c c}
\hline\hline
Source of Uncertainty & Energy Scale & Flux & Index & Curvature\\
\hline
MC shower interactions			& --	& 1\%	&	--	& -- \\
MC atmosphere simulation		& 7\%	&	&	--	&  -- \\
\hline
Instrument simulation / calibration	& 10\%	& 10\%	& --		& -- \\
Broken pixels				& --	& 5\%	& --		& -- \\
Live Time				& --	& $<5\%$ & --		& -- \\
\hline
Reconstruction and selection cuts	& 15\%	& 15\%	& 0.1  / 0.46	& 0.01 / 0.8\\
Background subtraction			& --	& 6\%/10\%	& 0.14 / 0.46	& 0.12 / 0.6 \\
\hline
Total					& 19\%	& 20\%/22\%	& 0.17 / 0.65	& 0.12 / 1.0\\
\hline\hline
\end{tabular}
\end{table*} 

The main sources of systematic uncertainties in the \phasetwo mono analysis presented in this publication, and their estimated contributions to the uncertainty on the spectral parameters, are summarised in Table \ref{SystErrorTable}. 
For each source of uncertainty the table gives the flux normalisation uncertainty,
the photon index uncertainty and the uncertainty on the curvature parameter $\beta$ (for the log-parabola model).
In addition, the energy scale uncertainty is given in the second column.
The energy scale uncertainty implies an additional uncertainty on the flux normalisation which depends on the steepness of the spectrum.
It is also relevant for the determination of the position of spectral features such as the SED maximum or EBL cutoff.
The procedures used here for estimating the systematic uncertainties generally repeat the procedures used for \hess I \citepads{2006A&A...457..899A}.
We highlight that the discussion in this section focuses specifically on the sources and analysis presented. A more general discussion of the systematic uncertainties of the \phasetwo mono analysis will be part of a future publication.

Except for background subtraction, all sources of uncertainty listed in Table~\ref{SystErrorTable} are related to the conversion of the measured event counts into flux.
This conversion is done using the instrument response functions (IRF) which are determined from Monte Carlo simulations.
The IRF uncertainties show how well the real instrument, after all calibrations, is described by the simulation.

The first group of uncertainties is related to the interaction of particles and their
production and to the absorption of Cherenkov light in the atmosphere.
The estimated uncertainty due to the shower interaction model does not exceed 1\% (for photon-induced showers).
The atmospheric uncertainties include the effects of the atmospheric density profile (which affects the height of shower maximum and Cherenkov light production) and the atmospheric transparency (light attenuation by Mie and Rayleigh scattering).
These effects were studied extensively during \hess phase I \citepads{2000APh....12..255B,2006A&A...457..899A,2014APh....54...25H}.
The uncertainties were found to be dominated by the atmospheric transparency, which has direct influence on the amount of Cherenkov light detected by the telescopes, thus affecting the energy reconstruction.
Data from the telescope radiometers and other atmospheric monitoring devices, as well as trigger rate data, are used to ensure good atmospheric conditions during the observations used in the analysis (see section~\ref{data_quality}).
For zenith angles relevant to this work, the remaining uncertainty on the absolute energy scale due to the atmosphere
is estimated to be $\approx \, 7$\% \citepads[similar to the uncertainty level reported in]{2006A&A...457..899A}.

The instrument simulation and calibration uncertainty includes all remaining instrumental effects, such as mirror reflectivity and electronics response.
These effects are controlled using various calibration devices \citepads{2004APh....22..109A},
as well as Cherenkov light from atmospheric muons \citepads{2003ICRC....5.2895L}.
The non-operational pixels in the CT5 camera ($< 5$\%) and the electronics dead time ($< 5$\%) contribute only marginally to the overall uncertainty.

The event reconstruction and selection uncertainties are derived from a comparison of the measured spectra with the results obtained using an alternative analysis chain (see Sect.~\ref{sect:crosscheck}).

Irregularities in the camera acceptance (e.g. due to non-operational pixels)
and the night sky background (e.g. bright stars)
can both have an effect on background subtraction.
The background subtraction errors are controlled in this study by
visually examining the raw and acceptance-corrected skymaps
(to ensure that there are no artefacts, e.g. from bad calibration of individual data runs),
as well as using additional dedicated tests and run quality selection.
As shown already in Sect.~\ref{section:results}, the width of the skymap
significance distributions is dominated by statistical errors.
This is ensured for both objects, \pks and \pg,
and throughout the entire energy range covered by this study
(see Figs.~\ref{pks2155_ebins} and \ref{pg1553_ebins}).
Hence, arguably, the effect of the background subtraction errors 
should not exceed the statistical uncertainties.
Consequently, the statistical uncertainties on the spectral parameters represent
a reasonably conservative estimate of the background subtraction uncertainties.
It should be noted, however, that the reflected-region background method, which is used 
for the spectral measurements,
is potentially more sensitive to non-axially symmetric effects in the camera acceptance than the ring background maps
(which use a 2D acceptance model).
We have investigated this further by splitting the full data set into two subsets,
one of which groups the data from runs taken with a wobble offset in right ascension (in either positive or negative direction)
and another one for the remaining runs (with wobble in declination).
The signal-to-background (S/B) ratios obtained with these subsets were compared to the full dataset S/B ratio.
It was found that the S/B ratio varied by $\approx 3$\%,
which is about twice the background subtraction accuracy observed
with the ring background method ($\approx 1.5$\% of the background level).
Therefore in Table \ref{SystErrorTable} the statistical uncertainties are doubled
to obtain the values for the background subtraction uncertainties.

The net effect of all uncertainties summed in quadrature is given in the last row of Table~\ref{SystErrorTable}.
It can be noted that the spectral index and curvature uncertainties are dominated
by the reconstruction, event selection and background subtraction uncertainties,
while the description of the atmosphere and instrument calibration contribute substantially 
to the energy scale and flux normalisation uncertainties.

It should lastly be highlighted that the systematic uncertainties are energy-dependent. In particular, the background subtraction uncertainties tend to become more important towards low energies, where the signal-to-background ratio is usually smaller.
For an analysis aiming at the lowest energies
this can lead to a large uncertainty in the measurement of spectral index and curvature,
especially for soft spectrum sources, as is the case for \pg.

In the context of variability studies, the uncertainty values presented in Table~\ref{SystErrorTable} can be considered as a conservative upper bound.
Preliminary studies of steady sources with \phasetwo suggest that the RMS variability induced by systematic effects is about 15-20\%,
a result similar to that for \hess I \citepads{2006A&A...457..899A}.
This suggests that at least some of the spectral measurement uncertainties are constant in time and could therefore be reduced by means of additional calibrations.
Variations related to changes in the atmosphere transparency can also be reduced by means of additional corrections \citepads{2014APh....54...25H}.

\section{Discussion}
\label{discussion}

The successful \phasetwo mono observations and analysis of \pks and \pg convincingly
demonstrate that the low energy part of the VHE spectrum is accessible
to the \hess experiment, following the addition of the CT5 instrument. This fact
makes EBL studies of high redshift AGNs by \phasetwo mono feasible, without the need
for strong theoretical biases on the intrinsic spectra or the need to rely on
spectral extrapolations using results from other instruments.

Here we consider EBL deabsorbed fits to the \phasetwo mono and contemporaneous
\fermi spectra for both AGNs. Our aim here is twofold. The first is to
investigate evidence for curvature in the two AGN intrinsic spectra, correcting
for EBL absorption effects. Second, given 
the present systematic uncertainties derived for these
data sets, we determine the corresponding uncertainties on the combined fit parameters.
Such considerations provide insight into the
constraining power of these results, under the assumption of both a specific EBL
model \citepads[in this work the one of][]{2008A&A...487..837F} and simple
underlying spectral shape.

The spectra in the \phasetwo mono energy range have been reconstructed with a spectral model corrected for EBL absorption. Furthermore, for \pg, whose redshift is
not well-constrained, we adopt the well-motivated value of $z = 0.49$
\citepads{2015ApJ...802...65A}.

\begin{table*}
\caption{Parameters obtained for the combined fit of the \fermi and H.E.S.S. data. The reference energy $E_0$ used here is 100 GeV. For both blazars, the log-parabola fits values are provided. For \pg, the values for the power-law model, which was marginally disfavoured, are also given. The last column gives the significance, obtained by comparing the $\chi^2$ values for the log-parabola model against those for the power-law model, using only statistical errors in the analysis.}
\label{table:Fit} 
\centering 
\begin{tabular}{c c c c c c c} 

\hline\hline 
Source &$ \phi_0 [10^{-11}\ {\rm cm}^{-2} {\rm s}^{-1}]$  &  $ \Gamma$ & $\beta$ &$log_{10}(E_{\rm peak} [{\rm GeV}])$& Sig. ($\sigma$) \\
\hline 
\pks&$ 2.35  \pm 0.10_{\rm stat}\pm 0.57_{\rm sys}$  & $2.30  \pm 0.04_{\rm stat}\pm 0.09_{\rm sys}$ &  $0.15  \pm 0.02_{\rm stat}\pm 0.02_{\rm sys}$ & $0.99 \pm 0.19_{\rm stat}\pm 0.19_{\rm sys}$& 5.1 \\
\pg&$ 5.97\pm 0.25_{\rm stat}\pm 2.19_{\rm sys}$  &  $1.68\pm 0.05_{\rm stat}\pm 0.13_{\rm sys}$ & --  & -- & --\\
\pg&$ 6.66\pm 0.42_{\rm stat}\pm 1.43_{\rm sys}$  &  $1.83\pm 0.08_{\rm stat}\pm 0.29_{\rm sys}$ & $0.12\pm 0.05_{\rm stat}\pm 0.13_{\rm sys}$  & $2.76 \pm 0.45_{\rm stat}\pm 0.93_{\rm sys}$& 2.2 \\

\hline 
\end{tabular}
\end{table*}

In order to look for a possible turnover in the intrinsic spectrum and, if present, to locate
the peak emission in the energy flux ($E^{2}{\rm d}N/{\rm d}E$) representation, the EBL deabsorbed 
\fermi and \phasetwo mono data points were fitted both separately and as a combined data set with 
power-law, broken power-law and log-parabola models. In the combined fit procedure, a consideration
of the systematic uncertainties for each of the data sets was taken into account in the analysis. 

For the \hess systematic uncertainties, the effect of the energy systematic uncertainty on 
the deabsorbed spectrum fit results was found to be the dominant contributing systematic.
The contribution of this uncertainty on the results was estimated through the shifting 
of the data points in the $E\,dN/dE$ representation by an energy scale factor of 19\% 
(see Table~\ref{SystErrorTable}) before applying the EBL deabsorbtion. The variation in 
the best-fit model, introduced via the application of this procedure within the full energy uncertainty
range, was then taken as the systematic contribution to the uncertainty on each model parameter (see
Table~\ref{table:Fit}).
An estimate of the size of the \fermi systematic uncertainties was also obtained, using the 
effective area systematic uncertainty, derived by the LAT collaboration\footnote{see
\url{http://fermi.gsfc.nasa.gov/ssc/data/analysis/scitools/Aeff_Systematics.html}.}. These
uncertainties were noted to be small in comparison to the statistical errors such that their further 
consideration could be safely neglected.

\begin{figure}
  \centering
\includegraphics[width=1.0\linewidth]{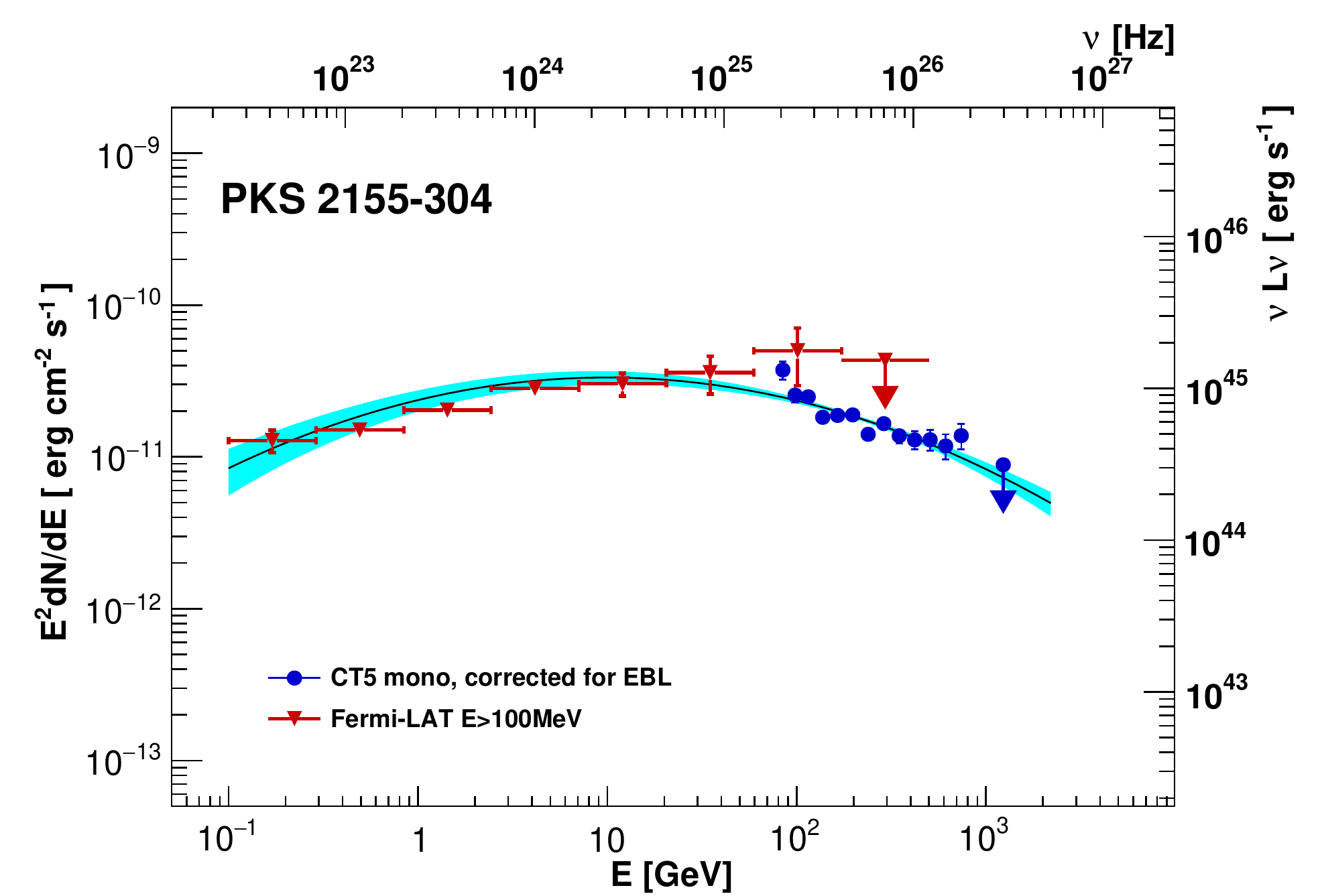}
  \caption{The energy spectrum of \pks obtained from the \phasetwo mono analysis (blue) of the 2013 data corrected for EBL absorption in comparison with the contemporaneous \fermi data with a minimal energy of 0.1~GeV (red). The black line is the best-fit log-parabola model to the points and the cyan butterfly indicates the 1~$\sigma$ region using only the statistical errors in the combined data set analysis. The right-hand y-axis shows the equivalent isotropic luminosity (not beaming corrected).}
  \label{pks2155_spectrum_ebl}
\end{figure}

In the case of \pks, separate fits of the \fermi and \phasetwo mono EBL deabsorbed data, 
the power-law model was found to provide a sufficient description in both cases.
The power-law fit of the \phasetwo mono 2013 data obtained an intrinsic 
spectral index of $\Gamma=2.49\pm 0.05$. Such an index appears somewhat softer than 
the power-law analysis of the \fermi contemporaneous data 
($\Gamma=1.82\pm 0.03$ see Table~\ref{FermiTab}). The spectral fits found for the 
combined data sets, dominated by the low energy data points where EBL effects can 
be neglected, allowed the continuity of the source spectrum to be probed. The fit of 
the combined \fermi and \phasetwo mono data with a log-parabola model was preferred at the
5.1~$\sigma$ level with respect to the power-law model (See Figure~\ref{pks2155_spectrum_ebl}). 
The broken power-law does not significantly improve the fit in this case. The results of the fit are
given in Table~\ref{table:Fit}. The peak flux position within the SED was at 
a moderate energy (around 10~GeV), in agreement with its 4-year averaged position found in 
the 3FGL.

\begin{figure}
  \centering
\includegraphics[width=1.0\linewidth]{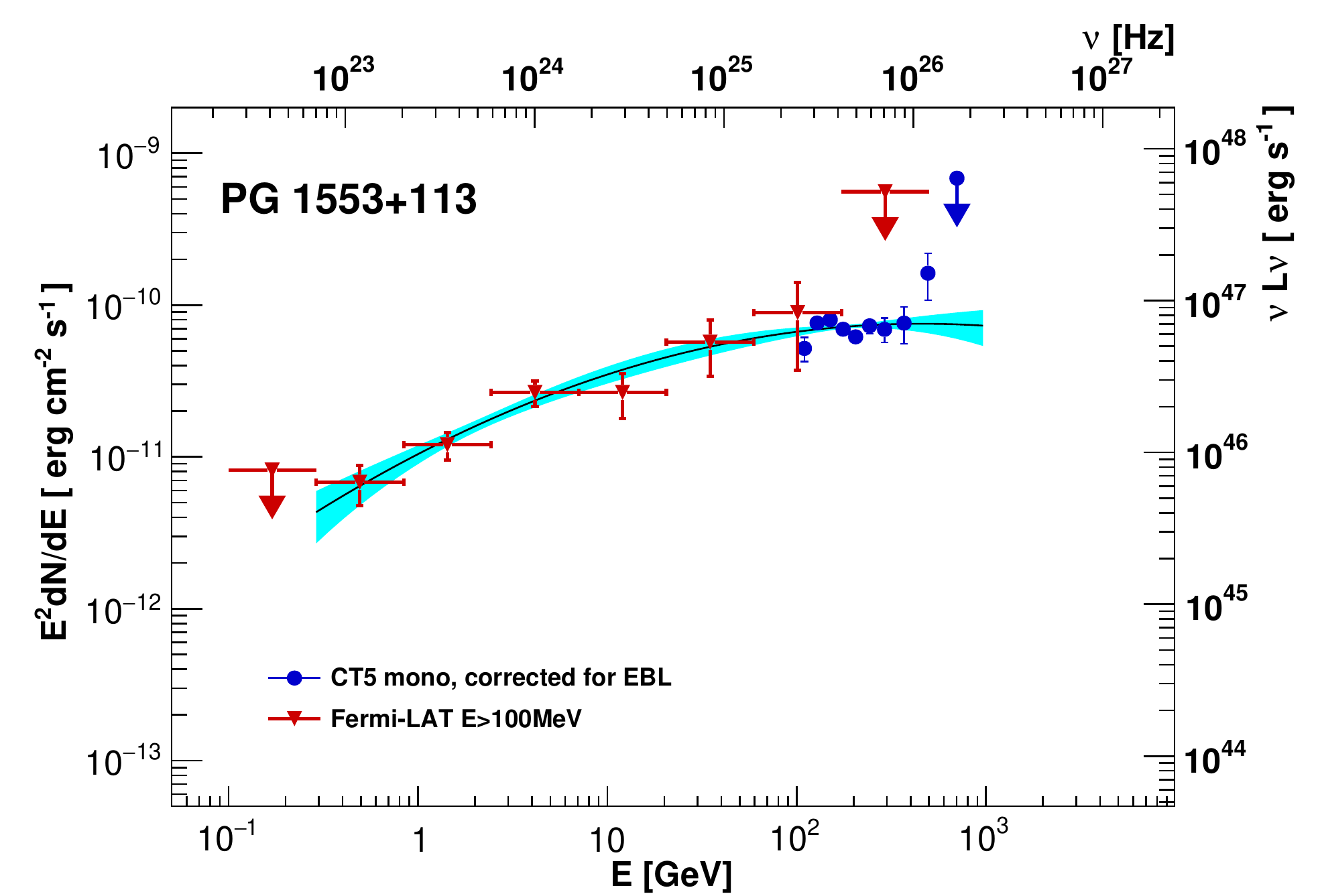}
  \caption{The energy spectrum of \pg obtained from the \phasetwo mono analysis (blue) corrected for EBL absorption in comparison with the contemporaneous \fermi data with a minimal energy of 0.1~GeV (red). The assumed redshift is z=0.49. The black line is the best-fit log-parabola model fit to the points and the cyan butterfly indicates the 1~$\sigma$ (statistical error only) uncertainty region. The right-hand y-axis shows the equivalent isotropic luminosity (not beaming corrected).}
  \label{pg1553_spectrum_ebl}
\end{figure}

For \pg, an EBL absorbed power-law fit to the \phasetwo mono spectra
required an intrinsic spectral index of $\Gamma=1.91\pm 0.13$. For comparison, 
Table~\ref{FermiTab} shows that the \fermi
spectral fits for power laws with thresholds of 100~MeV and 10~GeV 
give consistent spectral indices to this value. On the other hand, the 
fit of the combined \fermi and \phasetwo mono gamma-ray data, however, found a 
log-parabola model preferred at the 2.2~$\sigma$ level over the power-law model (See Figure~\ref{pg1553_spectrum_ebl}).
The fit values for these two spectral models are provided in Table~\ref{table:Fit}. 
The parameters that results from fits with a broken power-law being close to one of the single power-law model case.
The sizeable systematic errors, once also taken into account, however, weaken 
this preference. Thus, this only marginal improvement, brought by the log-parabola model, suggests that the observed 
softening of the \pg spectrum is predominantly introduced by VHE interaction on 
the EBL, a result consistent with that from other instruments which have searched 
for intrinsic curvature in the source's spectra \citepads{2015MNRAS.450.4399A}. 
 Furthermore, the constraint on the intrinsic peak position, at a value of 
0.6$^{+1.0}_{-0.4}$~TeV,
also carries significant uncertainties. 
This limitation is primarily due to the relatively small intrinsic curvature, 
limited lever arm (energy range coverage by the measurements), and the very soft 
observed spectral index in the \phasetwo mono band, which amplifies the effect of 
the energy scale uncertainty. This could be improved in the future via more accurate
calibration of the \phasetwo mono energy scale, using bright flaring or stable sources to
compare flux measurements with those of \fermi contemporaneous measurements as e.g in 
\citetads{2010A&A...523A...2M}.

In summary, the high-energy SED of \pg, corrected for EBL with the 
model of \citetads{2008A&A...487..837F}, assuming a redshift of 0.49, 
reveals only marginal evidence for intrinsic curvature once systematic 
uncertainties are taken into account. This result is compatible with a 
scenario in which the observed spectral downturn at an energy of around 
100 GeV is introduced through the attenuation at the highest energies 
is due to the interaction of VHE photons with the EBL. Contrary to 
this, in the case of \pks, the EBL corrected SED is better described by 
a log-parabola model than by a power-law. The addition of intrinsic 
spectral curvature or break is required to account for the data 
presented. Such a feature is naturally expected rather generically on 
physical grounds in the high energy region of the particle spectrum for 
both stochastic and shock acceleration mechanisms 
\citepads{1995ApJ...446..699P,1987MNRAS.225..335H,1998A&A...333..452K}.

\section{Conclusions}
\label{Conclusions}

Here we report, for the first time, \phasetwo mono blazar results following 
observations of \pks in 2013 and 2014 and \pg in 2013, taken with the new CT5 
instrument in monoscopic configuration. The successful analysis of 
these observations resulted in the detection of 
these two AGNs at levels of $\sim$42$~\sigma$ (36$~\sigma$) 
and $\sim$27$~\sigma$ (21$~\sigma$), respectively. 
For these results, low-energy thresholds of 80~GeV and 110~GeV, respectively, 
were achieved. These thresholds amount to a reduction by a factor of 2--3 
relative to that achieved in the CT1-4 cross-check results presented (see Figures  \ref{pks2155_spectrum} and   \ref{pg1553_spectrum}).
Furthermore, we note that the energy threshold achieved by the present 
\phasetwo mono analysis
remains limited by the accuracy of the background subtraction method,
rather than by the instrument trigger threshold.

Namely, at energies below the respective thresholds achieved for the \pks and
\pg datasets, the systematic uncertainties in background subtraction become
larger than the statistical uncertainties. The energy at which the transition
from statistics-dominated to systematics-dominated regime occurs depends on the
accuracy of background subtraction and the size of the dataset being analysed.
For the present analysis the level of systematic uncertainties in background
subtraction was found to be $\approx$ 1.5\% (for skymaps), which corresponds to
a minimal requirement for the signal-to-background ratio of $S/B > 7.5\%$ for a
$5\sigma$ detection (assuming normally distributed errors). This limitation does
not apply to the special case of gamma-ray pulsars, where the pulsar phasogram
can be used to define ``off regions'' for background subtraction.
Subsequent improvements and reduction in the energy threshold are likely to occur in the future.

A comparison of the emission level of \pks and \pg with their historic observations 
revealed both to be in low states of activity, with \pks found to be within
1~$\sigma$ of its mean quiescent level, as defined in \citetads{2010A&A...520A..83H}, 
during the 2013 \phasetwo observations. 
Temporal analysis of its emission during the campaign revealed mild 
($\sim$ 50\%) variability in the lightcurve of \pks between the 2013 and 2014 
\phasetwo data sets. No significant variability was found in the emission of \pg.
Further analysis of the \pks data, separating the two consecutive years 
of observations, revealed an enhancement in the flux state, by a 
factor of $\sim$60\%, in the 2014 data. Interestingly, a similar size increase in 
the flux level between the 2013 and 2014 fluxes is seen in the contemporaneous 
\fermi data (see Fig.~\ref{pks2155_SED_2013_2014}). Spectral analysis of the fluxes 
from these two different brightness periods, however, reveals no evidence for significant 
alteration of the spectral shape from either the \phasetwo mono or \fermi
observations. The change in source state between these periods 
therefore appears to be associated with a broad increase in the
source brightness in the 0.1--1000~GeV energy range.

Multi-wavelength SED plots containing the new \phasetwo data points for these observations of 
\pks and \pg, and their comparison with contemporaneous \fermi observations, are shown in 
Figs.~\ref{pks2155_spectrum},~\ref{pks2155_SED_2013_2014}, and~\ref{pg1553_spectrum}. 
Spectral analysis of the \phasetwo mono data indicate that
a log-parabola fit is preferred over a simple power-law or a broken power-law fit in 
both cases. The measurement of the curvature parameter in these fits, however, is marginal 
for \pg once the systematic errors are taken into account. Within their multi-wavelength SEDs, 
the presence of a strong spectral downturn feature, at an energy of $\sim 100$~GeV, is
apparent in both cases, consistent with previous multi-wavelength observations
made of these objects during low activity states 
\citepads{2009ApJ...696L.150A,2014A&A...571A..39H,2010ApJ...708.1310A,2012ApJ...748...46A,2015ApJ...799....7A}.
The introduction of such a feature at these
energies is expected through gamma-ray absorption on the EBL during their transit through 
extragalactic space. Adopting a specific EBL model, spectral fitting of the data, deabsorbed
on the EBL, indicates the presence of significant curvature in the intrinsic source spectrum for 
\pks, with the peak of the intrinsic SED sitting at an energy of $\sim 10$~GeV.
A similar EBL deabsorbed analysis for \pg reveals a milder level of curvature in the
intrinsic spectrum, suggesting that the peak of the intrinsic SED sits at an energy of $\sim 500$~GeV.
However, once systematic errors are taken into account, the intrinsic spectrum of \pg was found 
to be consistent with no curvature.
It therefore remains possible that the observed softening in the \pg spectra is purely introduced 
by VHE interaction on the EBL, and is not intrinsic to the source.

Our results demonstrate for the first time the successful 
employment of the monoscopic data from the new \phasetwo instrument (CT5) 
for blazar and other AGN studies.
These results mark a significant step forward in lowering the gamma-ray energy 
range that AGN may be probed in the \phasetwo era.
This reduction in the energy threshold opens up the opportunity to probe new
low-energy aspects about AGN fluxes, their variability, and their 
attenuation on the EBL out to larger redshifts than that probed 
previously in the \hess~I era.
Furthermore, coupled with the level of significance obtained for the detection of
both AGNs, the reduction in threshold offers great potential for temporally
resolving AGN lightcurves down to unprecedented temporal scales during flaring
episodes.

\begin{acknowledgements}
  The support of the Namibian authorities and of the University of Namibia in facilitating the construction and operation of H.E.S.S. is gratefully acknowledged, as is the support by the German Ministry for Education and Research (BMBF), the Max Planck Society, the German Research Foundation (DFG), the French Ministry for Research, the CNRS-IN2P3 and the Astroparticle Interdisciplinary Programme of the CNRS, the U.K. Science and Technology Facilities Council (STFC), the IPNP of the Charles University, the Czech Science Foundation, the Polish Ministry of Science and Higher Education, the South African Department of Science and Technology and National Research Foundation, the University of Namibia, the Innsbruck University, the Austrian Science Fund (FWF), and the Austrian Federal Ministry for Science, Research and Economy, and by the University of Adelaide and the Australian Research Council. We appreciate the excellent work of the technical support staff in Berlin, Durham, Hamburg, Heidelberg, Palaiseau, Paris, Saclay, and in Namibia in the construction and operation of the equipment. This work benefited from services provided by the H.E.S.S. Virtual Organisation, supported by the national resource providers of the EGI Federation.

  The \textit{Fermi}-LAT Collaboration acknowledges generous ongoing support from a number of agencies and institutes that have supported both the development and the operation of the LAT as well as scientific data analysis. These include the National Aeronautics and Space Administration and the Department of Energy in the United States, the Commissariat \`a l'Energie Atomique and the Centre National de la Recherche Scientifique / Institut National de Physique Nucl\'eaire et de Physique des Particules in France, the Agenzia Spaziale Italiana and the Istituto Nazionale di Fisica Nucleare in Italy, the Ministry of Education, Culture, Sports, Science and Technology (MEXT), High Energy Accelerator Research Organization (KEK) and Japan Aerospace Exploration Agency (JAXA) in Japan, and the K.~A.~Wallenberg Foundation, the Swedish Research Council and the Swedish National Space Board in Sweden.
  
  Additional support for science analysis during the operations phase is gratefully acknowledged from the Istituto Nazionale di Astrofisica in Italy and the Centre National d'\'Etudes Spatiales in France.

  This research has made use of NASA's Astrophysics Data System.
  This research has made use of the SIMBAD database, operated at CDS, Strasbourg, France.
  This research made use of Enrico, a community-developed Python package to simplify Fermi-LAT analysis \citepads{2013arXiv1307.4534S}.
\end{acknowledgements}

\bibliography{HESS2_AGN}

\end{document}